\documentclass[12pt]{iopart}
\pdfoutput=1
\expandafter\let\csname equation*\endcsname\relax
\expandafter\let\csname endequation*\endcsname\relax

\usepackage{color}
\usepackage[breakwords]{truncate}
\usepackage[font={small}]{caption}
\usepackage{graphicx}
\usepackage{cite} 
\usepackage{subcaption}
\usepackage{etoolbox}
\usepackage[section]{placeins}
\usepackage[sectionbib]{chapterbib}
\usepackage{amsfonts}
\usepackage{amsmath}
\usepackage[bookmarks,bookmarksopen,bookmarksdepth=2]{hyperref}
\usepackage{bm}
\makeatletter
\newrobustcmd{\fixappendix}{%
  \patchcmd{\l@section}{1.5em}{7em}{}{}%
  \patchcmd{\l@subsection}{2.3em}{7em}{}{}%
}
\makeatother

\appto\appendix{
\addtocontents{toc}{\fixappendix}
\addtocontents{toc}{\protect\setcounter{tocdepth}{1}}}

\newcommand{\bea}{\begin{eqnarray}}
\newcommand{\eea}{\end{eqnarray}}

\begin{document}
\newcommand{\ben}[1]{\textcolor{blue}{\textbf{#1}}}

\title{Generating stochastic trajectories with global dynamical constraints}
\author{Benjamin De Bruyne}
\address{LPTMS, CNRS, Univ.\ Paris-Sud, Universit\'e Paris-Saclay, 91405 Orsay, France}
\author{Satya N. Majumdar}
\address{LPTMS, CNRS, Univ.\ Paris-Sud, Universit\'e Paris-Saclay, 91405 Orsay, France}
\author{Henri Orland}
\address{Institut de Physique Théorique, Université
Paris-Saclay, CEA, 91191 Gif/Yvette Cedex, France}
\author{Gr{\'e}gory Schehr }
\address{Sorbonne Universit\'e, Laboratoire de Physique Th\'eorique et Hautes Energies, CNRS UMR 7589, 4 Place Jussieu, 75252 Paris Cedex 05, France}
\eads{\mailto{benjamin.debruyne@centraliens.net}, \mailto{satya.majumdar@universite-paris-saclay.fr},
\mailto{henri.orland@ipht.fr},
\mailto{gregory.schehr@u-psud.fr}}
\begin{abstract}
We propose a method to exactly generate Brownian paths $x_c(t)$ that are constrained to return to the origin at some future time $t_f$, with a given fixed area $A_f = \int_0^{t_f}dt\, x_c(t)$ under their trajectory. We derive an exact effective Langevin equation with an effective force that accounts for the constraint. In addition, we develop the corresponding approach for discrete-time random walks, with arbitrary jump distributions including L\'evy flights, for which we obtain an effective jump distribution that encodes the constraint. Finally, we generalise our method to other types of dynamical constraints such as a fixed occupation time on the positive axis $T_f=\int_0^{t_f}dt\, \Theta\left[x_c(t)\right]$ or a fixed generalised quadratic area $\mathcal{A}_f=\int_0^{t_f}dt \,x_c^2(t)$. 
\end{abstract}

\section{Introduction}
Brownian motion is the basis of many applications in science. In one dimension, a Brownian motion evolves according to the Langevin equation
\begin{align}
  \dot x(t) = \sqrt{2\,D}\,\eta(t)\,,\label{eq:eom}
\end{align}
where $D$ is the diffusion coefficient and $\eta(t)$ is an uncorrelated Gaussian white noise with zero mean and the correlator $\langle \eta(t)\eta(t')\rangle=\delta(t-t')$. For many applications, it is necessary to simulate Brownian paths numerically. This can be easily done by discretizing the Langevin equation (\ref{eq:eom}) over small time increments $\Delta t$:
\begin{align}
  x(t+\Delta t) = x(t) +  \sqrt{2D}\,\eta(t)\,\Delta t\,,\label{eq:discL}
\end{align}
and drawing independently at each step a Gaussian random increment $\sqrt{2D}\,\eta(t)\,\Delta t$ of zero mean and variance $2D\Delta t$. In some applications, one is only interested in paths that satisfy given constraints. For instance, if one is simulating the displacements of foraging animals close to their home range, a natural constraint on the trajectories is that the animals must return to their nest after a fixed amount of time \cite{Giuggioli05,Randon09,MajumdarCom10,Murphy92,Boyle09}. Such constrained motion is usually referred to as a bridge as the initial and final points are fixed. Other notable constrained processes are the excursions, meanders, reflected motions, etc \cite{Yor2000,Majumdar05Ein,MP2010,Dev2010,PY2018}. These processes have wide applications in the context of behavioral ecology \cite{Giuggioli05,Randon09,MajumdarCom10,Murphy92,Boyle09}, financial stock markets \cite{Shepp79,Majumdar08}, or in statistical testing \cite{CB2012,Kol1933}. 

A natural question that arises is how to generate these constrained trajectories. A naive solution would be to generate free paths and discard the ones that do not satisfy the constraint. Unfortunately, this method turns out to be computationally wasteful as the trajectories satisfying the constraint are typically rare \cite{BCDG2002,GKP2006,GKLT2011,KGGW2018,Gar2018,Rose21,Rose21area,CLV21} and therefore difficult to obtain. Fortunately, for the case of Brownian motion, there exist several efficient methods, based on the so-called Doob transform \cite{Doob,Pitman}, to generate particular types of constrained trajectories. One of them, which is quite versatile, consists in writing an effective Langevin equation with an effective force that implicitly accounts for the constraint. For instance, to generate a Brownian bridge $x_B(t)$ of duration $t_f$ with the bridge constraint $x_B(0)=x_B(t_f)$, the effective Langevin equation reads \cite{MajumdarEff15,CT2013}
\begin{align}
  \dot x_B(t) =   - \frac{x_B(t)}{t_f-t} + \sqrt{2\,D}\,\eta(t)\,,\label{eq:eff}
\end{align}
where the subscript $B$ refers to ``bridge'' and the first term in the right-hand side is the effective force that accounts for the bridge constraint. Simulating Brownian bridges can then be easily done by discretizing the effective Langevin equation over small time increments. Effective Langevin equations have been obtained for several constrained processes such as excursions, meanders  \cite{CT2013,MajumdarEff15,Orland,Baldassarri2021} and more recently for interacting particles, such as non-intersecting Brownian bridges \cite{Grela2021}. It was also recently shown that this concept can be applied to discrete-time random walks with arbitrary jump distributions, including fat-tailed ones \cite{DebruyneRW21}, as well as to non-Markovian processes, such as the run-and-tumble motion \cite{DebruyneRTP21}.

While the effective Langevin equation has proven to be a successful technique to generate constrained paths with local constraints, such as the initial and final points in the Brownian bridge, an effective Langevin equation is still lacking for global constraints, such as constraints on time-integrated quantities. One prominent example of a time-integrated quantity for Brownian motion is the area $A(t)$ under its trajectory
\begin{align}
  A(t) = \int_0^t d T\, x( T)\,.\label{eq:Ac}
\end{align}
An interesting question to ask is: ``How to generate Brownian paths that return to the origin after a fixed amount of time $t_f$ with a fixed area $A_f$ under their trajectory?''. One of the goals of this paper is to provide an answer to this question. The area under a Brownian motion has received sustained interest and attention in various fields such as mathematics \cite{Takacs91,Janson07} and computer science due to its relation to algorithmic problems
\cite{KnuthThe98,FlajoletOn98,MajumdarExact02,Majumdar05Ein}. In physics, the area under a Brownian motion plays a central role in many problems, including 
$(1+1)$-dimensional fluctuating interfaces that we discuss below.

An extensively studied model of $(1+1)$-dimensional fluctuating interfaces is governed by the celebrated Kardar-Parisi-Zhang (KPZ) equation which, in its simplest form, describes the spatio-temporal evolution of a height function $H(x,t)$ of an interface on a linear substrate of length $L$ \cite{KPZ}:
\begin{align}
  \partial_t H(x,t) = \partial^2_{x} H(x,t) + \lambda \left(\partial_x H(x,t)\right)^2 + \xi(x,t)\,,\label{eq:KPZ}
\end{align}
where $\xi(x,t)$ is a Gaussian white noise of zero mean with a correlator $\langle \xi(x,t)\xi(x',t') \rangle=\delta(x-x')\delta(t-t')$. When the non-linear term is absent ($\lambda=0$), the KPZ equation reduces to the well-known Edwards-Wilkinson interface model \cite{EW}. On a substrate of finite size $L$, the KPZ equation displays two regimes: (i) a growing regime for time $t \ll L^z$ (where the dynamical exponent is $z=3/2$) and (ii) a stationary regime when $t \gg L^z$ \cite{HHZ95,Krug97b}. While there have been extensive recent studies on the growing regime, which is connected to random matrix theory \cite{KK10,Cor12,HHT15}, here our focus is on the stationary regime. For $t \gg L^z$, 
the joint distribution of the heights $\{ H(x,t)\}$, for $0 \leq x \leq L$, does not reach a time-independent stationary state, since the mean height $\overline{H(x,t)}=\frac{1}{L}\int_0^L dx\, H(x,t)$ keeps growing with time. However, if one defines the relative heights as 
\begin{align}
  h(x,t) = H(x,t) - \overline{H(x,t)}\,, \label{eq:relh}
\end{align}
then the joint distribution of the relative heights $h(x,t)$ for $t \gg L^z$ does reach a stationary state. For periodic boundary conditions $h(0)=h(L)$, this stationary distribution is given by \cite{Majumdar04Flu1,Majumdar05Flu2}
\bea \label{stat_pdf}
P_{\rm stat}[\{h(x)\}] = \frac{1}{Z_L} e^{- \frac{1}{2} \int_0^L \left[\partial_x h(x)\right]^2 \, dx}\; \delta(h(0)-h(L)) \; \delta\left(\int_0^L h(x)\, dx\right) \;.
\eea
While the first delta-function represents the periodic boundary conditions, the second one reflects the constraint satisfied by the relative heights in (\ref{eq:relh}). 
Indeed, the definition in (\ref{eq:relh}) imposes the {\it global} constraint that the total area under the relative heights is exactly zero. In (\ref{stat_pdf}), $Z_L$
is the partition function that normalises the probability measure. Note that this stationary measure (\ref{stat_pdf}) holds both for the KPZ as well as the EW interface ($\lambda=0$). With the identification $h(x) \to x(t)$ and $x \in [0,L]$ transposing to $t \in [0,t_f]$ with $t_f=L$, the 
stationary measure in (\ref{stat_pdf}) corresponds to a Brownian bridge ($x(0) = x(t_f)$) with the global constraint that the area under the bridge is exactly zero. 
To sample the distribution $P_{\rm stat}[\{h(x)\}]$ in (\ref{stat_pdf}), one then needs to generate Brownian bridges constrained by the zero area condition. 
This is a concrete physical example of a Brownian bridge with a global constraint. This global contraint played a crucial role on the behavior of many stationary observables, such
as on the distribution of the maximal relative height \cite{Majumdar04Flu1,Majumdar05Flu2} and on the spatial persistence \cite{MD2006}. The effect of this global zero area constraint on the relative heights was also studied in various generalisations of interfaces with a non-Brownian stationary measure \cite{Majumdar01,GMOR2007}.

Another generalisation of the stationary measure of the Brownian interface with a zero area constraint (\ref{stat_pdf}) corresponds to studying $(1+1)$-dimensional solid-on-solid models 
on a discrete lattice of size $L$ with periodic boundary conditions of the form \cite{Schehr10}
\bea \label{stat_pdf_disc}
P_{\rm stat}[\{h_i\}] = \frac{1}{Z_L} e^{- K \sum_{i=1}^L |h_{i+1} - h_i|^\alpha}\;\delta(h_{L+1}-h_1) \;\delta\left( \sum_{i=1}^L h_i\right) \;,
\eea
where $h_i$ represents the stationary height of the interface at site $i$ and $\alpha > 0$. Here, instead of a Brownian motion in space, the interface height in the
stationary state, performs a random walk in space
\bea \label{rw_hi}
h_{i+1} = h_i + \eta_i \;,
\eea 
where $\eta_i$'s are independent and identically distributed (IID) random noises, each drawn from a PDF $f(\eta) \propto e^{-K\,|\eta|^\alpha}$. More generally, the stationary
measure reads \cite{Schehr10}
\bea \label{stat_pdf_disc_gen}
P_{\rm stat}[\{h_i\}] = \frac{1}{Z_L} \left[\prod_{i=1}^L f(h_{i+1} - h_i)\right]\;\delta(h_{L+1}-h_1) \;\delta\left( \sum_{i=1}^L h_i\right) \;,
\eea
where $f(\eta)$ may have a fat tail corresponding to a L\'evy interface in space. In these discrete cases also, to sample the stationary measure (\ref{stat_pdf_disc_gen}), one needs
to generate discrete time random walk bridges with the global zero area constraint, upon the identification $h_i \to x_i$ and the space index $i$ in the interface model identified with the time step of the random walk bridge. This is then a discrete-time random walk analogue of its continuous-time counterpart, namely the Brownian bridge in the presence of the zero area constraint.

In this paper, we derive exactly an effective Langevin equation to generate Brownian bridges with a fixed arbitrary area under their trajectory. 
We show that in order to do this, one needs to keep track of both the position and the area of the process as a function of the evolving time. We will see that this
joint process of the position and the area is Markovian, which allows us to write down an effective Langevin equation for the joint process.   
We then show how this method can be generalised to the case of discrete-time random walks, with arbitrary jump distributions, for which we derive an effective jump distribution that accounts for the constraint. To illustrate our method, we apply it to the case of random walks with Gaussian or Cauchy jump distributions. Finally, we show how the effective Langevin equation can be generalised to other constraints, such as a fixed occupation time on the positive axis or a fixed quadratic area under the trajectory. 

The remaining of this paper is organized as follows. In Section \ref{sec:BM}, we outline the derivation of the effective Langevin equation to generate bridge Brownian motion with a fixed area under its trajectory. In Section \ref{sec:rw}, we develop the corresponding discrete-time random walk approach to generate bridge trajectories with a fixed area. We obtain an effective jump distribution which we implement numerically for the case of Gaussian and Cauchy random walks. In Section \ref{sec:gen}, we discuss a generalisation of our method to generate Brownian bridges with a fixed occupation time on the positive axis or a fixed quadratic area under its trajectory. Finally, in Section \ref{sec:ccl}, we conclude with a summary and perspectives for further research. Some detailed calculations are presented in Appendices. 

\section{Generating Brownian bridges with a fixed area}\label{sec:BM}

We now consider a Brownian motion up to some fixed time $t_f$ and with a total fixed area under the curve $A_f = \int_0^{t_f} x_c(t) \, dt$ where 
$x_c(t)$ denotes the position of the Brownian motion at some intermediate time $t$ and the subscript $c$ refers to the fact that the motion is ``constrained''. To proceed, it is convenient to define a dynamical area variable $A_c(t) = \int_0^t x_c( T) \, d T$. Now we consider the process $(x_c(t), A_c(t))$ jointly whose evolution is governed locally in time by the Markov dynamics 
\begin{subequations}
\begin{align}
  \dot x_c(t) &= \sqrt{2D}\,\eta(t)\,,\\
  \dot A_c(t) &= x_c(t)\,,
\end{align}
\label{eq:evxA}
\end{subequations}
\hspace*{-0.4cm}  The constraint on the trajectory is that it must start and return to the origin after a fixed amount of time $t_f$ with a given area $A_f$ under its trajectory, namely
\begin{align}
  x_c(0)=x_c(t_f)=0\,,\quad A_c(t_f)=A_f\,.\label{eq:constc}
\end{align}
Thus we can think of this joint process as a bridge in the plane, going from the initial value $(x_c(0)=0, A_c(0)=0)$ to the final value $(x_c(t_f)=0, A_c(t_f)=A_f)$.  The derivation of the effective Langevin equation for this joint process, in the presence of the bridge constraint (\ref{eq:constc}), then closely follows the derivation for a one-dimensional bridge in  
\cite{MajumdarEff15}.

\begin{figure}[t]
  \begin{center}
    \includegraphics[width=0.6\textwidth]{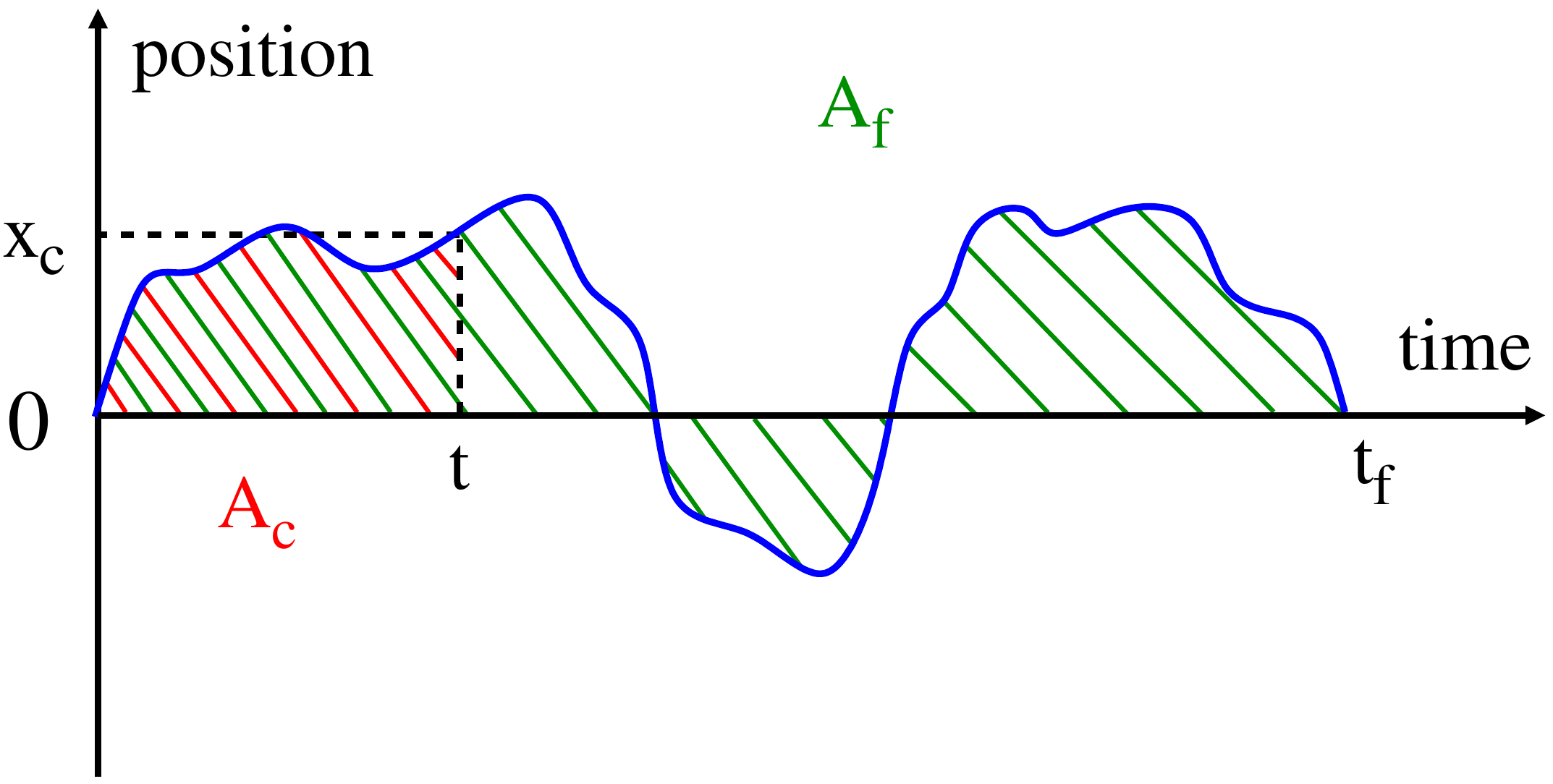}
    \caption{Schematic representation of a bridge Brownian path of duration $t_f$ with a total area $A_f$ under its trajectory. By using the Markov property, the trajectory can be decomposed into two independent paths: one in the time interval $[0,t]$ where the particle propagates from the origin to $x_c$ with an area $A_c$, another one in the time interval $[t,t_f]$ where the particle propagates from $x_c$ with an initial area $A_c$ to the origin with an area $A_f$. }
    \label{fig:bridge}
  \end{center}
\end{figure}

To derive an effective Langevin equation for this joint process, we first write the constrained joint probability distribution $P_c(x_c,A_c,t\,|\,A_f,t_f)$ for the position $x_c$  and the area $A_c$ at an intermediate time $t<t_f$. Due to the Markov property (see figure \ref{fig:bridge}), it can be written as the product
\begin{align}
  P_c(x_c,A_c,t\,|\,A_f,t_f) = \mathcal{N} P(x_c,A_c,t\,|\,0,0,0) \,P(0,A_f,t_f\,|\,x_c,A_c,t)\,,\label{eq:PcN}
\end{align}
where $P(x_c,A_c,t\,|\,0,0,0)$ is the propagator of the \emph{free} Brownian motion (without constraints), i.e. the probability distribution that the particle reaches $x_c$ at a time $t$ with an area $A_c$ under its trajectory given that it started initially at $t=0$ at the origin with zero initial area. By integrating the constrained propagator over $x_c$ and $A_c$, and by using Chapman-Kolmogorov property of transition probabilities, the normalisation constant is found to be $\mathcal{N}^{-1}=P(0,A_f,t_f\,|\,0,0,0)$. For conciseness, we introduce the notations 
\begin{align}
  P(x_c,A_c,t\,|\,0,0,0) \equiv P(x_c,A_c,t)\,,\quad P(0,A_f,t_f\,|\,x_c,A_c,t) \equiv \tilde Q(x_c,A_c,t)\,,\label{eq:defPQ}
\end{align}
where we also used the fact that the process is time translational invariant, which means that the propagator depends only on the time difference between the final and initial time.
In (\ref{eq:defPQ}) we have also suppressed the explicit dependence of $\tilde Q$ on $A_f$ -- actually we will shortly show that the evolution equation for $\tilde Q$ does not contain $A_f$ explicitly and the dependence of $\tilde Q$ on $A_f$ only appears through its initial condition. With these notations, the constrained propagator (\ref{eq:PcN}) reads
\begin{align}
  P_c(x_c,A_c,t\,|\,A_f,t_f) = \frac{P(x_c,A_c,t)\, \tilde Q(x_c,A_c,t)}{P(0,A_f,t_f)}\label{eq:Pc}
\end{align}
The propagator $P(x,A,t)$ denotes the probability density that the joint process arrives at $(x,A)$ at time $t$ starting from $(0,0)$. In contrast, the propagator $\tilde Q(x,A,t)$, by the second relation in (\ref{eq:defPQ}), denotes the 
probability density that the joint process starting at $(x,A)$ at time $t$, arrives at $(0,A_f)$ at time $t_f$. Below, we first derive the evolution equations for $P$ and $\tilde Q$ separately and, subsequently, for their product. 

To derive the evolution equation for $P$, we consider evolving the joint process from $t$ to $t+dt$. In this interval, we see from the equation of motion (\ref{eq:eom})-(\ref{eq:Ac}) that the particle traveled from $x-\sqrt{2\,D}\,\eta(t)$ to $x$ and that the area changed from $A-x\,dt$ to $A$. Averaging over all possible noise realizations $\eta(t)$, we find
\begin{align}
  P(x,A,t+dt) = \langle P(x-\sqrt{2\,D}\,\eta(t)\,dt,A-x\,dt,t)\rangle_{\eta(t)}\,. \label{eq:Pder}
\end{align}
Taking the limit $dt\rightarrow 0$ gives
\begin{align}
\partial_t P(x,A,t) = D \partial^2_{x} P(x,A,t) -x\, \partial_A P(x,A,t)\,,\label{eq:Peq}
\end{align}
which must be solved with the initial condition $P(x,A,t=0)=\delta(x)\delta(A)$.

One can similarly derive the equation for $\tilde Q$ as follows. We again evolve the joint process from $t$ to $t+dt$. 
In this small time interval $dt$, the position has moved to $x + \sqrt{2 D} \eta(t)\, dt$ while the area has become $A + x(t) \,dt$. For the subsequent evolution from $t + dt$ to $t_f$, the "new initial value" of the process is $(x + \sqrt{2 D} \eta(t)\, dt, A + x(t) \,dt)$.
Hence, averaging over all possible values of $\eta(t)$, we find 
\begin{align}
 \tilde Q(x,A,t) = \langle \tilde Q(x+\sqrt{2\,D}\,\eta(t)\,dt,A+x\,dt,t+\delta t)\rangle_\eta\,. \label{eq:Qeq}
\end{align}
Taking the limit $dt\rightarrow 0$ gives
\begin{align}
-\partial_{t} \tilde Q(x,A,t) = D \partial^2_{x} \tilde Q(x,A,t) +x \,\partial_A \tilde Q(x,A,t)\,. \label{eq:back}
\end{align}
This equation evolves from $t$ to $t_f$ and the condition at $t=t_f$ is $\tilde Q(x,A,t=t_f)=\delta(x)\delta(A-A_f)$. To ease notations,
using the time translational invariance of the process, we define 
\bea \label{def_Q}
\tilde Q(x,A,t) = Q(x,A,t_f-t) \;.
\eea
Therefore $Q(x,A,t)$ evolves via
\bea \label{FP_Q}
\partial_{t} Q(x,A,t) = D \partial^2_{x} Q(x,A,t) +x \,\partial_A Q(x,A,t) \;,
\eea
with the initial condition $Q(x,A,t=0) = \delta(x)\delta(A-A_f)$ which comes from the condition on $\tilde Q$ at $t_f$. Therefore the joint propagator for the bridge in (\ref{eq:Pc}) reads
\begin{align}
  P_c(x_c,A_c,t\,|\,A_f,t_f) = \frac{P(x_c,A_c,t)\, Q(x_c,A_c,t_f-t)}{P(0,A_f,t)}\label{eq:Pcbis} \;,
\end{align}
where $P$ and $Q$ evolve respectively via (\ref{eq:Peq}) and (\ref{FP_Q}).

\begin{figure}[t]
\subfloat{%
 \includegraphics[width=0.5\textwidth]{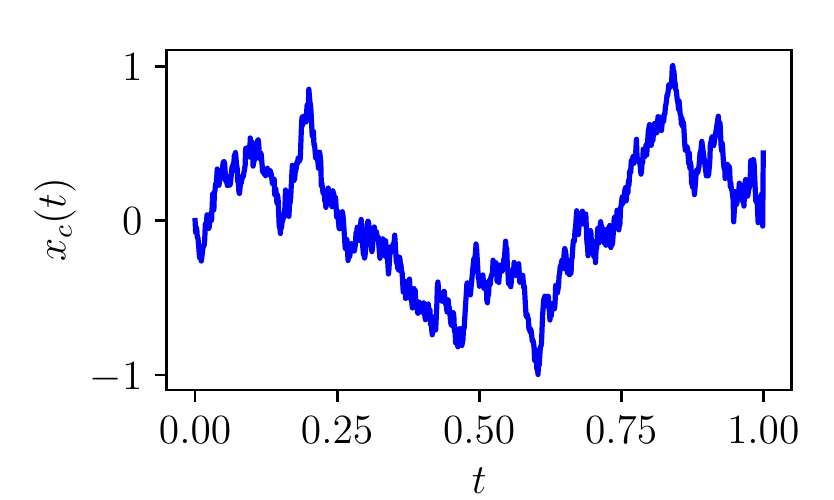}%
}\hfill
\subfloat{%
  \includegraphics[width=0.5\textwidth]{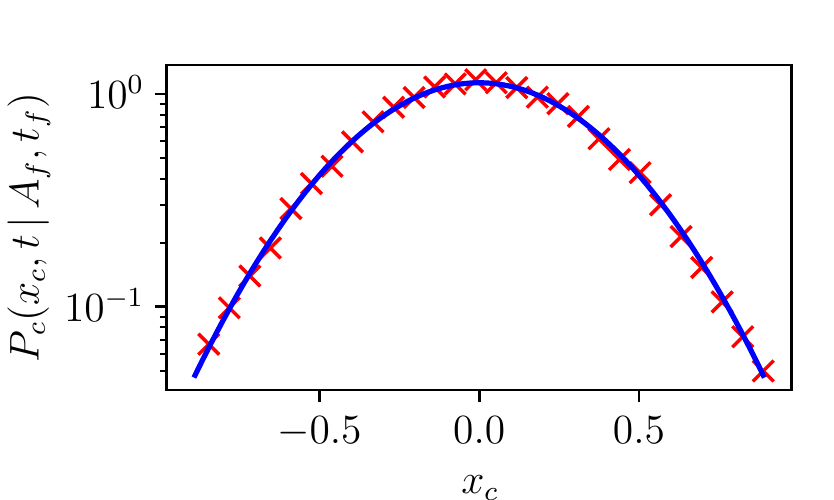}%
}\hfill\\
\subfloat{%
 \includegraphics[width=0.5\textwidth]{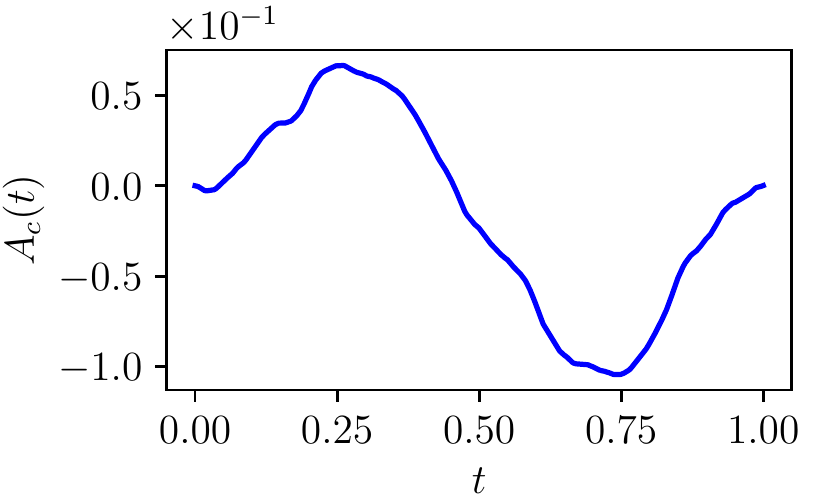}%
}\hfill
\subfloat{%
  \includegraphics[width=0.5\textwidth]{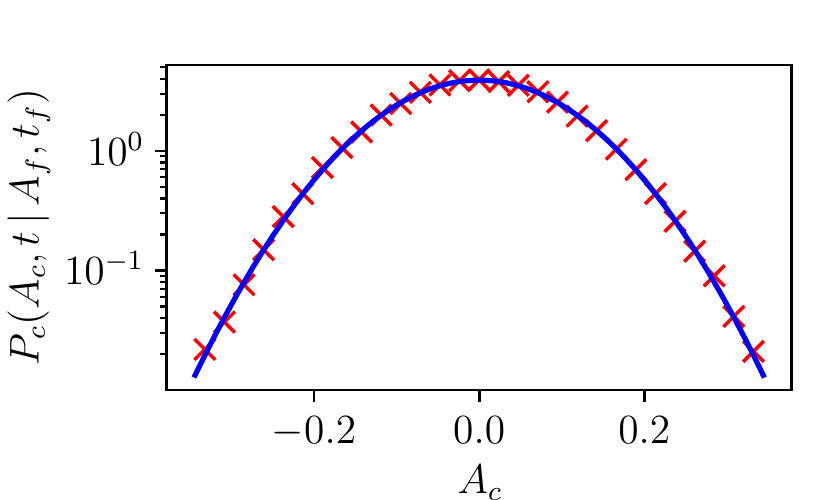}%
}\hfill
\caption{\textbf{Left panel:} A typical trajectory $x_c(t)$ vs $t$ (top) and $A_c(t)$ vs $t$ (bottom) for a bridge Brownian motion of duration $t_f=1$ with a zero area constraint $A_f=0$ generated by the effective Langevin equation (\ref{eq:effLBM}) for $D=1$. \textbf{Right panel:} Marginal position (top) and area (bottom) distributions at $t=t_f/2$ for a bridge Brownian motion of duration $t_f=1$ with a zero area constraint $A_f=0$. These marginal distributions, obtained numerically by sampling the trajectories from the effective Langevin equation (\ref{eq:effLBM}), are compared with the theoretical predictions in (\ref{marg_xc_BM}) and (\ref{marg_Ac_BM}). The numerical correlation is $\langle (x_c-\mu_x) (A_c-\mu_A) \rangle \approx-5\times 10^{-5}$ and the theoretical one is $0$ from (\ref{correlxA_cont}) for the choice $t = t_f/2$.}\label{fig:constrained}
\end{figure}

Our goal is to show that the constrained propagator (\ref{eq:Pcbis}) also satisfies a Fokker-Plank equation. By taking a time derivative of the constrained propagator (\ref{eq:Pc}) and using the equations (\ref{eq:Peq}) and (\ref{FP_Q}) satisfied by the free propagators, we find that the constrained propagator satisfies the Fokker-Plank equation
\begin{eqnarray}
\hspace*{-2cm}  \partial_t P_c(x_c,A_c,t) &=& D \partial_{x_c}\left[\partial_{x_c} P_c(x_c,A_c,t)-2\,P_c(x_c,A_c,t)\,\partial_{x_c} \ln(Q(x_c,A_c,t_f-t))\right] \nonumber \\
\hspace*{-2cm}  &-&x_c \partial_{A_c} P_c(x_c,A_c,t)\,,\label{eq:Pceq}
\end{eqnarray}
where we have omitted the conditional dependence in $P_c(x_c,A_c,t\,|\,A_f,t_f)$ for conciseness. The equation (\ref{eq:Pceq}) is very similar to the one satisfied by the free propagator (\ref{eq:Peq}), except that it has an additional force term $2D\,\partial_{x_c} \ln(Q(x_c,A_c,t_f-t))$. Therefore, bridge trajectories with a fixed area can be generated using the effective Langevin equations
\begin{subequations}
\begin{align}
  \dot x_c(t) &= \sqrt{2\,D}\,\eta(t) + 2\,D\,\partial_{x_c}\ln[Q(x_c(t),A_c(t),t_f-t)]\,,\\
  \dot A_c(t) &= x_c(t)\,,
\end{align}
\label{eq:effL}
\end{subequations}
where $Q(x_c,A_c,t)$ is the solution of the Fokker-Plank equation (\ref{eq:back}) with the initial condition $Q(x_c,A_c,0)=\delta(x_c)\delta(A_c-A_f)$. The solution of this equation is given by (see \ref{app:area}):
\begin{align}
  Q(x_c,A_c,t) = \frac{\sqrt{3}}{2\pi Dt^2}\,\exp\left[-\frac{1}{D}\left(\frac{3\,(A_c-A_f)\,(A_c-A_f+x_c\,t)}{t^3}+\frac{x_c^2}{t}\right)\right]\,.\label{eq:Qs}
\end{align}
One can check that the propagator is indeed normalised $\int_{-\infty}^{\infty} dx_c\int_{-\infty}^{\infty}dA_c\, Q(x_c,A_c,t)=1$. Inserting the expression into the effective Langevin equations (\ref{eq:effL}), we find
\begin{subequations}
\begin{align}
  \dot x_c(t) &= \sqrt{2\,D}\,\eta(t) - \frac{6\,(A_c(t)-A_f)}{(t_f-t)^2} -\frac{4\,x_c(t)}{t_f-t}\,,\label{eq:effLBMa} \\
    \dot A_c(t) &= x_c(t)\,. \label{eq:effLBMb}
\end{align}
  \label{eq:effLBM}
\end{subequations}
This effective Langevin equation is the generalisation of equation (\ref{eq:eff}) presented in the introduction, with the additional area constraint. By discretizing it over small time increments, it can be used to generate constrained trajectories (see left panel in figure \ref{fig:constrained}). 
In the right panel in figure \ref{fig:constrained}, we computed numerically the marginal probability distributions of the position and the area at some intermediate time $t=t_f/2$, by generating trajectories from (\ref{eq:effLBM}). This is compared to the theoretical marginal distributions of the position and area for the bridge, which can be easily computed by substituting the free propagators $Q(x_c,A_c,t)$ and $P(x_c,A_c,t)=Q(x_c,A_f-A_c,t)$ from (\ref{eq:Qs}) in (\ref{eq:Pc}), which gives
\begin{align}
 P_c(x_c,A_c,t\,|\,A_f,t_f) =  \frac{1}{2\pi\,\sqrt{\det(\bm{\Sigma})}}\,e^{-\frac{1}{2}\,(\mathbf{x_c}-\bm{\mu})^T \bm{\Sigma}^{-1} \,(\mathbf{x_c}-\bm{\mu})}\,,\label{eq:Pcbza}
\end{align}
where $\mathbf{x_c}=(x_c,A_c)$. The parameters of this bi-variate Gaussian distribution, namely the vector $\bm{\mu}$ and the $2 \times 2$ connected correlation matrix $\bm{\Sigma}$,  are given by
\bea
&&\bm{\mu}=\left(\mu_x=\frac{6A_f t(t_f-t)}{t_f^3},\,\mu_A = \frac{A_f t^2 (2t-3t_f)}{t_f^3}\right) \label{mu_vector_cont}\\
&&\bm{\Sigma} = \begin{pmatrix}  
\langle (x_c - \mu_x)^2 \rangle & \langle (x_c -\mu_x)(A_c - \mu_A) \rangle \\[1em]
\langle (A_c -\mu_A)(x_c - \mu_x) \rangle & \langle (A_c -\mu_A)^2 \rangle  \label{sigma_matrix_cont}
\end{pmatrix} \;,
\eea
where
\bea 
&& \sigma_x^2 = \langle (x_c - \mu_x)^2\rangle = D\,\frac{2  t (t_f-t) \left(3 t^2-3 t\,
   t_f+t_f^2\right)}{t_f^3} \;, \label{correlx_cont}\\
&& \sigma_A^2 =\langle (A_c -\mu_A)^2 \rangle =    D\,\frac{2 t^3 (t_f-t)^3}{3 t_f^3}\;,  \label{correlA_cont} \\
&& \langle (A_c -\mu_A)(x_c - \mu_x)\rangle  = D\, \frac{ t^2 (t_f-t)^2 (t_f-2 t)}{t_f^3} \;. \label{correlxA_cont}
\eea
Note that the dependence on $A_f$ appears only in the average vector $\bm{\mu}$, but not in the correlation matrix $\bm{\Sigma}$. The marginal distributions can then be obtained by integrating (\ref{eq:Pcbza}) over $x_c$ or $A_c$ respectively and one gets, as expected, Gaussian distributions
\bea
&& P_c(x_c,t|A_f,t_f) =  \int_{-\infty}^\infty P_c(x_c,A_c,t|A_f,t_f)\, dA_c  =  \frac{1}{\sqrt{2\pi \sigma_x^2}}\,e^{-\frac{(x_c -\mu_x)^2}{2 \sigma_x^2}}\;, \label{marg_xc_BM} \\
&& P_c(A_c,t|A_f,t_f) =  \int_{-\infty}^\infty P_c(x_c,A_c,t|A_f,t_f)\, dx_c =   \frac{1}{\sqrt{2\pi \sigma_A^2}}\,e^{-\frac{(A_c -\mu_A)^2}{2 \sigma_A^2}}\;. \label{marg_Ac_BM} 
\eea
These marginal distributions are plotted in the right panel in figure \ref{fig:constrained} and compared to numerical simulations using the 
effective evolution equations (\ref{eq:effLBM}), finding excellent agreement. As an additional check, we compare the numerical correlation $\langle (A_c -\mu_A)(x_c - \mu_x) \rangle$ and the theoretical one, given in (\ref{correlxA_cont}), to probe the joint distribution of $x_c$ and $A_c$ beyond its marginal distributions.

\section{Generating discrete-time bridge random walks with a fixed area}\label{sec:rw}
In this section, we outline the derivation of the discrete-time counterpart of the effective Langevin method discussed in the previous section. The derivation closely follows \cite{DebruyneRW21}.
We consider a discrete-time random walk $x_n$ that evolves according to the Markov rule
\begin{subequations}
\begin{align}
  x_{m+1} = x_m + \eta_m\,, \label{eq:eomrw}
\end{align}
starting from $x_0=0$, where $\eta_m$' are independent and identically distributed (i.i.d.) random variables drawn from a normalised distribution $f(\eta)$. We define the dynamical area under the trajectory $A_m$ of the random walk after $m$ steps as
\begin{align}
  A_m = \sum_{i=0}^m x_i\,.\label{eq:ad}
\end{align} 
\label{eq:evolD}
\end{subequations}
\hspace*{-0.2cm}It is again convenient to consider jointly the variables $x_m$ and $A_m$. The constraints on the trajectories $x_{c,m}$ of duration $n$ with fixed area $A_f$ are:
\begin{align}
  x_{c,0}=x_{c,n}=0\,,\quad A_{c,n}=A_f\,.\label{eq:consd}
\end{align}
The constraints (\ref{eq:consd}) are to be seen as the discrete-time counterpart of the constraints (\ref{eq:constc}).
Analogously to the continuous-time constrained propagator (\ref{eq:Pcbis}), the discrete-time constrained propagator $P_c(x_c,A_c,m\,|\,A_f,n)$ for the position $x_c$ and the area $A_c$ at step $m$ for a discrete-time bridge random walk of length $n$ and total area $A_f$ is given by the normalised product
\begin{align}
P_c(x_c,A_c,m\,|\,A_f,n) =   \frac{P(x_c,A_c,m)\,Q(x_c,A_c,n-m)}{P(0,A_f,n)}\,,\label{eq:Pcd}
\end{align}
where the two propagators satisfy the recursive relations
\begin{subequations}
\begin{align}
  P(x,A,m) &= \int_{-\infty}^\infty d\eta\, P(x-\eta,A-x,m-1)\,f(\eta)\,,\label{eq:forwd}\\
  Q(x,A,m) &= \int_{-\infty}^\infty d\eta\, Q(x+\eta,A+x+\eta,m-1)\,f(\eta)\,,\label{eq:backd}
\end{align}
\label{eq:PQdis}
\end{subequations}
\hspace*{-0.2cm}with the initial conditions $P(x,A,0)=\delta(x)\delta(A)$ and $Q(x,A,0)=\delta(x)\delta(A-A_f)$.
Our goal is to show that the constrained propagator (\ref{eq:Pcd}) also satisfies a recursive relation. By using once the recursion (\ref{eq:backd}) on $Q(x_c,A_c,n-m)$ in (\ref{eq:Pcd}), we find that the constrained propagator satisfies the recursive equation
\begin{align}
  P_c(x_c,A_c,m\,|\,A_f,n) = \int_{-\infty}^\infty d\eta\, P_c(x_c-\eta,A_c-x_c,m-1\,|\,A_f,n)\,\tilde f(\eta\,|\,x_c,A_c,A_f,m,n)\,,\label{eq:pcd}
\end{align}
where the effective jump distribution is given by
\begin{align}
  \tilde f(\eta\,|\,x_c,A_c,A_f,m,n) = f(\eta)\,\frac{Q(x_c+\eta,A_c+x_c+\eta,n-m-1)}{Q(x_c,A_c,n-m)}\,.\label{eq:effJ}
\end{align}
This effective jump distribution is a generalisation of (21) in Ref.~\cite{DebruyneRW21}, with the additional area constraint. This distribution is parametrized by the current position $x_c$, current area $A_c$, final area $A_f$, current number of steps $m$ and the total length of the bridge $n$, which makes it non-stationary. One can obtain an explicit expression for the propagator in (\ref{eq:effJ}) as the recursive relations (\ref{eq:PQdis}) can be solved in Fourier space and give
\begin{subequations}
\begin{align}
P(x,A,m) &= \int_{-\infty}^{\infty}\int_{-\infty}^{\infty} \frac{dk}{2\pi} \,\frac{d\lambda}{2\pi} \,e^{-ikx-i\lambda A} \prod_{l=1}^{m} \,\left[\hat f(k+l\,\lambda)\right]\,,\\
  Q(x,A,m) &= \int_{-\infty}^{\infty}\int_{-\infty}^{\infty} \frac{dk}{2\pi} \,\frac{d\lambda}{2\pi} \,e^{-ikx-i\lambda (A-A_f)} \prod_{l=0}^{m-1} \,\left[\hat f(k-l\,\lambda)\right]\,,
\end{align}\label{eq:Qf}
\end{subequations}
where 
\begin{align}
  \hat f(k) = \int_{-\infty}^{\infty} d\eta \,f(\eta)\, e^{ik\eta}\,,\label{eq:hatf}
\end{align}
is the Fourier transform of the jump distribution. For some specific jump distributions, it is possible to find an exact analytical expression for the effective distribution (\ref{eq:effJ}). When an exact expression is difficult to obtain, or when no direct sampling methods exist, one can employ an acceptance-rejection sampling (ARS) algorithm (see e.g. \cite{Gilks92}). We briefly recall this algorithm here for completeness and refer the reader to \cite{DebruyneRW21} for a detailed discussion. 

For the ARS method to be applicable, the effective jump distribution should be bounded by the free jump distribution, i.e., we should be able to find a constant $c_{m,n}(x_c,A_c,A_f)\geq 1$ (independent of $\eta$), such that
\begin{align}
  \tilde f(\eta\,|\,x_c,A_c,A_f,m,n) \leq  c_{m,n}(x_c,A_c,A_f)\,f(\eta)\,,\quad \forall \eta\,.\label{eq:constant}
\end{align}
Under this condition, one can then sample realisations of the effective distribution by proceeding through the following steps
\begin{enumerate}
  \item Draw a candidate random number $\eta'$ from the free distribution $f(\eta')$\,,
  \item Accept the candidate $\eta'$ with probability $p_\text{accept}(\eta',x_c,A_c,A_f)$ given by
  \begin{align}
    p_\text{accept}(\eta',x_c,A_c,A_f,m,n) = \frac{ \tilde f(\eta'\,|\,x_c,A_c,A_f,m,n)}{c_{m,n}(x_c,A_c,A_f)\,f(\eta')}\,,\label{eq:paccept}
  \end{align}
  \item Reject the candidate otherwise and look for another one from step 1.
\end{enumerate}
In the next sections, we provide examples by applying our method to the case of random walks with a Gaussian or Cauchy jump distributions.

\subsection{Generating Gaussian bridge random walks with a fixed area}
We consider a Gaussian bridge random walk with a fixed total area $A_f$. The free jump distribution is given by
\begin{align}
  f(\eta) = \frac{1}{\sqrt{2\pi\sigma^2}}e^{-\frac{\eta^2}{2\sigma^2}}\,,\label{eq:fG}
\end{align}
where $\sigma^2$ is the variance of the jump distribution.
For this particular distribution, the free propagators (\ref{eq:Qf}) can be computed explicitly and are given by (see \ref{app:xAG})
\begin{subequations}
\begin{align}
P(x,A,m) &= \frac{\sqrt{3}}{\pi \,\sigma ^2\,
   m \sqrt{m^2-1}}\, \exp \left(-\frac{6 A^2-6 A (m+1) x+\left(2 m^2+3 m+1\right) x^2}{m \left(m^2-1\right) \sigma ^2}\right)\,,\label{eq:PRW}\\
  Q(x,A,m) = &\frac{\sqrt{3} \sqrt{2 m-1}}{\pi\,\sigma^2  \sqrt{m\left(m^2+m\right) \left(2 m^2-3 m+1\right)}  }\, \times \nonumber \\
  \quad&\exp \left(-\frac{6 (A-A_f)^2+6 (A-A_f) (m-1) x+\left(2 m^2-3 m+1\right) x^2}{m \left(m^2-1\right) \sigma ^2}\right)\,.\label{eq:QRW}
\end{align}
\label{eq:PQGRW}
\end{subequations}
One can check that these propagators are indeed normalised. The effective jump distribution (\ref{eq:effJ}) becomes another Gaussian distribution
\begin{align}
   \tilde f(\eta\,|\,x_c,A_c,A_f,m,n)  = \frac{1}{\sqrt{2\pi\sigma_{m,n}^2}}e^{-\frac{(\eta-\mu_{m,n})^2}{2\sigma_{m,n}^2}}\,,\label{eq:feffG}
\end{align}
where the mean and variance are now given by
\begin{subequations}
\begin{align}
  \mu_{m,n} &= -\frac{2 (3 (A_c-A_f)+2\, n\, x_c-2 \,m\, x_c-x_c)}{(1+n-m) (n-m)}\,,\\
  \sigma^2_{m,n} &= \sigma ^2\frac{ (n-m-1) (n-m-2)}{(n-m+1) (n-m)}\,.
\end{align}
\label{eq:musig}
\end{subequations}
The effective equations of the motion for the constrained random walk can therefore be written as
\begin{subequations}
\begin{align}
  x_{c,m} &= x_{c,m-1}  -\frac{2 (3 (A_c-A_f)+2\, n\, x_c-2 \,m\, x_c-x_c)}{(1+n-m) (n-m)} +  \sigma\,\frac{\sqrt{ (n-m-1) (n-m-2)}}{\sqrt{(n-m+1) (n-m)}} \,\eta_m\,,\\
   A_{c,m} &= A_{c,m-1} +  x_{c,m}\,,
\end{align}
\label{eq:effeqd}
\end{subequations}
\hspace*{-0.2cm}where $\eta_m$'s are i.i.d. random variables drawn from a Gaussian distribution with zero mean and unit variance. These effective equations generalise the one obtained in \cite{DebruyneRW21}, with the additional area constraint. Interestingly, we remark that the variance $\sigma_{m,n}^2$ in the second line of (\ref{eq:musig}) vanishes for $m=n-2$ and $m=n-1$. This means that the last two jumps of the walk are deterministic with value $\mu_{m,n}$ given in the first line in (\ref{eq:musig}).

\begin{figure}[t]
\subfloat{%
 \includegraphics[width=0.5\textwidth]{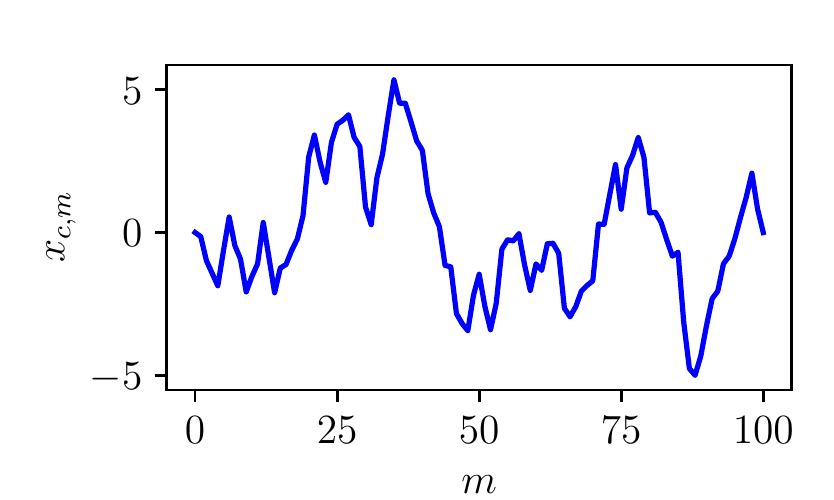}%
}\hfill
\subfloat{%
  \includegraphics[width=0.5\textwidth]{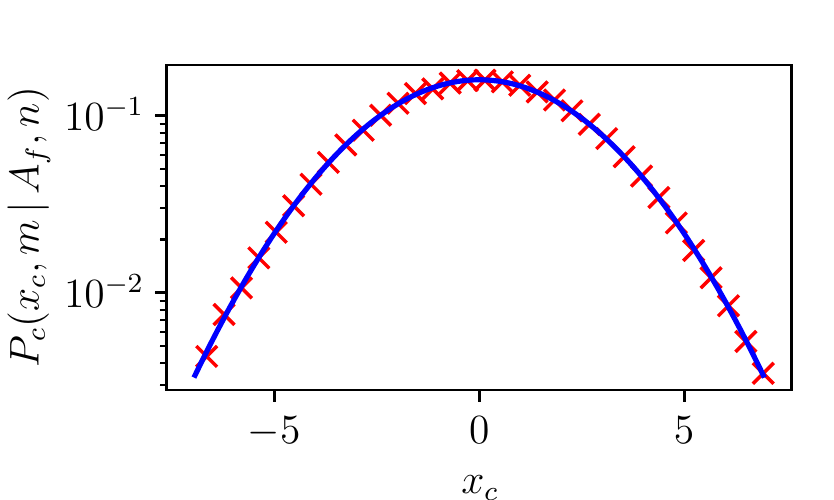}%
}\hfill\\
\subfloat{%
 \includegraphics[width=0.5\textwidth]{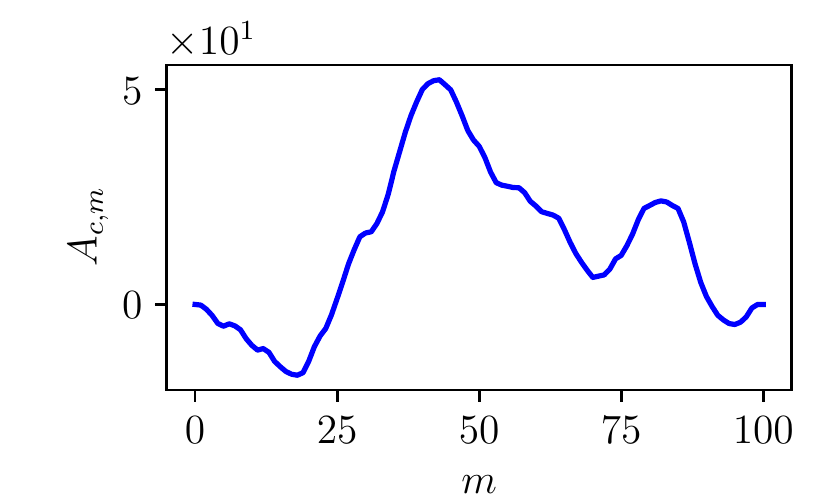}%
}\hfill
\subfloat{%
  \includegraphics[width=0.5\textwidth]{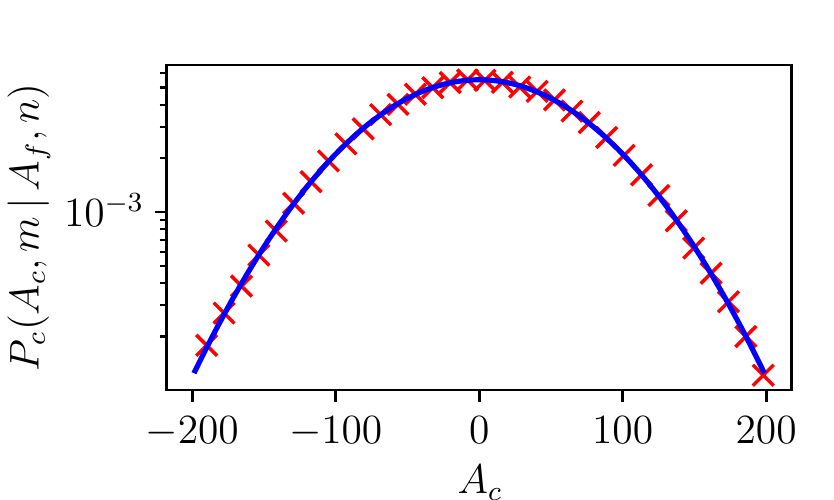}%
}\hfill
\caption{\textbf{Left panel:} A typical trajectory $x_{c,m}$ vs $m$ (top) and $A_{c,m}$ vs $m$ (bottom) of a Gaussian bridge random walk of $n=100$ steps with a zero area constraint $A_f=0$ generated by the effective jump distribution (\ref{eq:feffG}) with the choice $\sigma^2=1$. \textbf{Right panel:} Position (top) and area (bottom) marginal distributions at $m=50$ for a bridge random walk of $n=100$ steps with a zero area constraint. The distributions, obtained numerically by sampling the trajectories from the effective Langevin equation (\ref{eq:feffG}), are compared with the theoretical predictions in (\ref{marg_xc}) and (\ref{marg_Ac}). The connected correlator $\langle (x_c-\mu_x)(A_c-\mu_A)\rangle$ is measured numerically to be $\approx 3.15$, while the theoretical prediction 
in (\ref{correlxA}) for $m=50$, $n=100$ and $\sigma^2=1$ is given by 
$20825/6666 = 3.12406\ldots$. }\label{fig:GRWconstrained}
\end{figure}

It is interesting to verify that these discrete-time Langevin equations (\ref{eq:effeqd}) do converge to their continuous time counterparts in Eq.~(\ref{eq:effLBM}). 
To see this, we set $\sigma^2 = 2 D \Delta t$, $n = t_f/\Delta t$ and $m = t/\Delta t$ in (\ref{eq:effeqd}) and take the limit $\Delta t \to 0$, keeping $t, t_f$ and $D$ fixed.
This nicely reproduces the continuous time equations (\ref{eq:effLBM}) obtained in the previous section.
These effective equations can be used to generate constrained random walks (see left panel in figure \ref{fig:GRWconstrained}).
In the right panel in figure \ref{fig:GRWconstrained}, we computed numerically the marginal probability distributions of the position and the area at some intermediate time by generating bridge trajectories from (\ref{eq:feffG}). This is compared to the theoretical marginal distributions of the position and the area, which can be easily computed by substituting the free propagators (\ref{eq:PQGRW}) in (\ref{eq:Pcd}), which gives
\begin{align}
 P_c(x_c,A_c,m\,|\,A_f,n) =  \frac{1}{2\pi\,\sqrt{\det(\bm{\Sigma})}}\,e^{-\frac{1}{2}\,(\mathbf{x_c}-\bm{\mu})^T \bm{\Sigma}^{-1} \,(\mathbf{x_c}-\bm{\mu})}\,,\label{eq:PcG}
\end{align}
where $\mathbf{x_c}=(x_c,A_c)$. The parameters of this bi-variate Gaussian distribution, namely the vector $\bm{\mu}$ and the $2 \times 2$ connected correlation matrix $\bm{\Sigma}$,  are given by
\bea 
&&\bm{\mu}=\left(\mu_x=\frac{6A_f m(n-m)}{n(n^2-1)},\,\mu_A=\frac{A_f m(m+1)(3n-2m-1)}{n(n^2-1)}\right)  \label{mu_vector} \\
&&\bm{\Sigma} = \begin{pmatrix} 
\langle (x_c - \mu_x)^2 \rangle & \langle (x_c -\mu_x)(A_c - \mu_A) \rangle \\[1em]
\langle (A_c -\mu_A)(x_c - \mu_x) \rangle & \langle (A_c -\mu_A)^2 \rangle \label{sigma_matrix}
\end{pmatrix} \;,
\eea
where the three independent matrix elements are given by
\bea 
\hspace*{-2cm}&&\sigma_x^2 = \langle (x_c - \mu_x)^2 \rangle = \sigma^2\,\frac{m  (n-m) \left(3 m^2-3 m n+n^2-1\right)}{n \left(n^2-1\right)} \label{correlx} \\
\hspace*{-2cm}&&\sigma_A^2= \langle (A_c -\mu_A)^2\rangle = \sigma^2\, \frac{m (m+1)  (n-m) (n-m-1) [n+1-2 m (m-n+1)]}{6 n \left(n^2-1\right)} \label{correlA}\\
\hspace*{-2cm}&& \langle (x_c -\mu_x)(A_c - \mu_A) \rangle= \sigma^2 \, \frac{m (m+1)(n-m-1) (n+1-2 m) (n-m)}{2 n \left(n^2-1\right)} \label{correlxA} \;. 
\eea
Note that, as in the continuous time case, the dependence on $A_f$ appears only in the average vector $\bm{\mu}$, but not in the correlation matrix $\bm{\Sigma}$. The marginal distributions are simply Gaussians, obtained by integrating (\ref{eq:PcG}) over $x_c$ or $A_c$ respectively
\bea
&& P_c(x_c,m|A_f,n) =  \int_{-\infty}^\infty P_c(x_c,A_c,m|A_f,n)\, dA_c = \frac{1}{\sqrt{2\pi \sigma_x^2}}\,e^{-\frac{(x_c -\mu_x)^2}{2 \sigma_x^2}} \;, \label{marg_xc} \\
&& P_c(A_c,m|A_f,n) =  \int_{-\infty}^\infty P_c(x_c,A_c,m|A_f,n)\, dx_c  =  \frac{1}{\sqrt{2\pi \sigma_A^2}}\,e^{-\frac{(A_c -\mu_A)^2}{2 \sigma_A^2}}\;. \label{marg_Ac} 
\eea
These marginal distributions are plotted in the right panel in figure \ref{fig:GRWconstrained} and compared to numerical simulations using the 
effective evolution equations (\ref{eq:effeqd}), finding excellent agreement. As an additional check, we compare the numerical correlation $\langle (A_c -\mu_A)(x_c - \mu_x) \rangle$ and the theoretical one, given in (\ref{correlxA}), to probe the joint distribution of $x_c$ and $A_c$ beyond its marginal distributions.

\subsection{Generating Cauchy bridge random walks with a fixed area}

We now consider a discrete-time random walk in (\ref{eq:evolD}) where the jump distribution $f(\eta)$ does not have a finite variance.
These walks are usually called ``L\'evy flights''. An example is when the jumps have a Cauchy distribution
\begin{align}
    f(\eta) = \frac{1}{\gamma\,\pi}\,\frac{1}{\left[1+\left(\frac{\eta}{\gamma}\right)^2\right]}\,,\label{eq:cauchyDist}
\end{align}
where $\gamma$ is the scale of the distribution. Contrary to random walks with a finite variance jump distribution, the L\'evy walks do not converge to Brownian motion in the large $n$ limit. When the L\'evy walk is constrained to come back to a fixed final position $0$, this is called a L\'evy bridge and the distribution of the area under such a L\'evy bridge was studied in \cite{Schehr10}. Here we ask a different question, in the spirit of this paper: how do we generate such L\'evy bridges of $n$ steps with a fixed area $A_f$? For simplicity, we will focus here on L\'evy bridges with a Cauchy jump distribution given in (\ref{eq:cauchyDist}).  

The procedure is exactly similar to that in the previous section for Gaussian random walk bridges. The two propagators given in (\ref{eq:Qf}) for general jump distributions $f(\eta)$, 
read for the Cauchy case, 
\begin{subequations}
\begin{align}
P(x,A,m) &= \frac{1}{4\pi^2}\int_{-\infty}^{\infty}\int_{-\infty}^{\infty} dk \,d\lambda \,e^{-ikx-i\lambda A- \sum_{l=1}^{m} \,\gamma\, |k+l\,\lambda|}\,,\\
  Q(x,A,m) &= \frac{1}{4\pi^2}\int_{-\infty}^{\infty}\int_{-\infty}^{\infty} dk \,d\lambda \,e^{-ikx-i\lambda (A-A_f)-\sum_{l=0}^{m-1} \,\gamma\, |k-l\,\lambda|} \,.
\end{align}
\label{eq:PQcau}
\end{subequations}
where we used the fact that the Fourier transform of the Cauchy jump distribution is $\hat f(k)=\int_{-\infty}^{\infty} d\eta\, e^{ik\eta} f(\eta)=e^{-\gamma |k|}$.
Changing the cartesian variables $k,\lambda$ to polar variables $r,\theta$ and performing the integral over $r$ gives
\begin{align}
  P(x,A,m) &= \frac{1}{4\pi^2}\int_{0}^{2\pi}d\theta\, \frac{1}{[ix\sin(\theta)+iA\cos(\theta)+ \gamma\sum_{l=1}^{m} \, |\sin(\theta)+l\,\cos(\theta)|]^2}\,,\label{eq:PCf}\\
  Q(x,A,m) &= \frac{1}{4\pi^2}\int_{0}^{2\pi}d\theta\, \frac{1}{[ix\sin(\theta)+i(A-A_f)\cos(\theta)+ \gamma\sum_{l=0}^{m-1} \, |\sin(\theta)-l\,\cos(\theta)|]^2} \label{eq:PCb}\,.
\end{align}
The effective jump distribution (\ref{eq:effJ}) becomes
\begin{align}
  \tilde f(\eta\,|\,x_c,A_c,A_f,m,n) = f(\eta)\,\frac{Q(x_c+\eta,A_c+x_c+\eta,n-m-1)}{Q(x_c,A_c,n-m)}\,,\label{eq:feffC}
\end{align}
where $Q(x,A,m)$ (which does depend on $A_f$ also) is given in (\ref{eq:PCb}).

\begin{figure}[t]
\subfloat{%
 \includegraphics[width=0.5\textwidth]{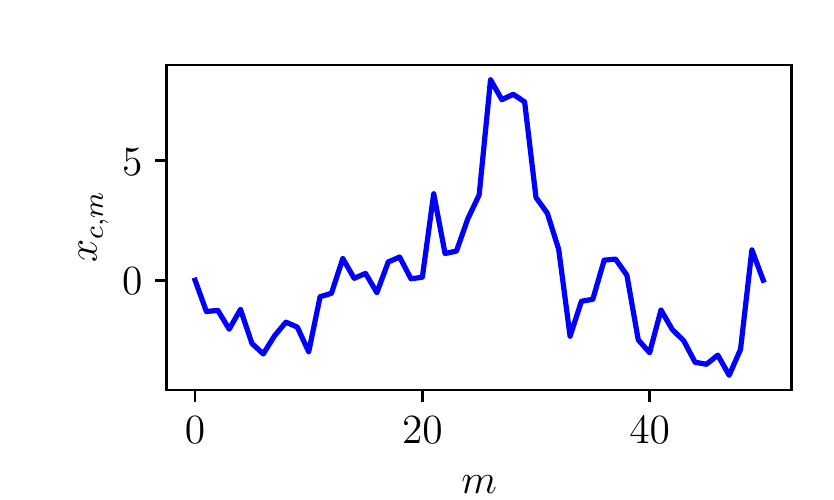}%
}\hfill
\subfloat{%
  \includegraphics[width=0.5\textwidth]{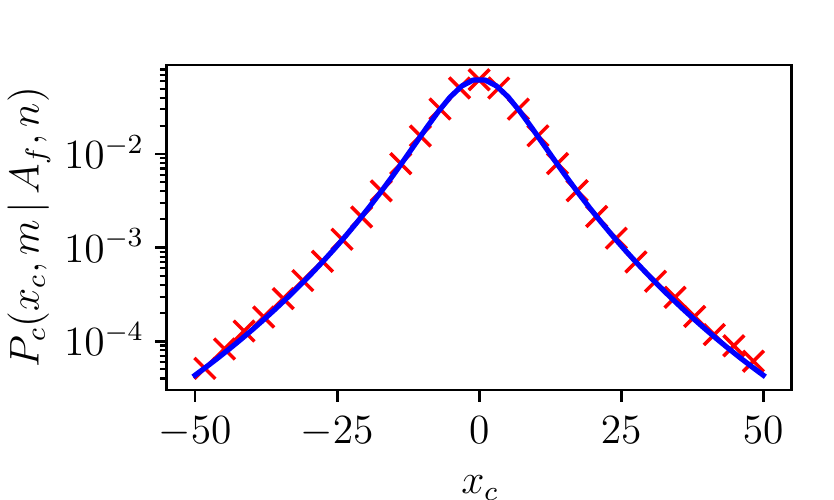}%
}\hfill\\
\subfloat{%
 \includegraphics[width=0.5\textwidth]{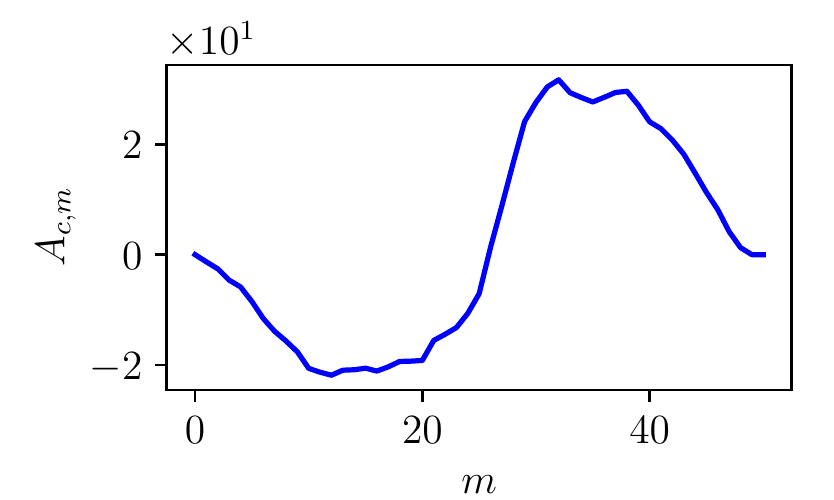}%
}\hfill
\subfloat{%
  \includegraphics[width=0.5\textwidth]{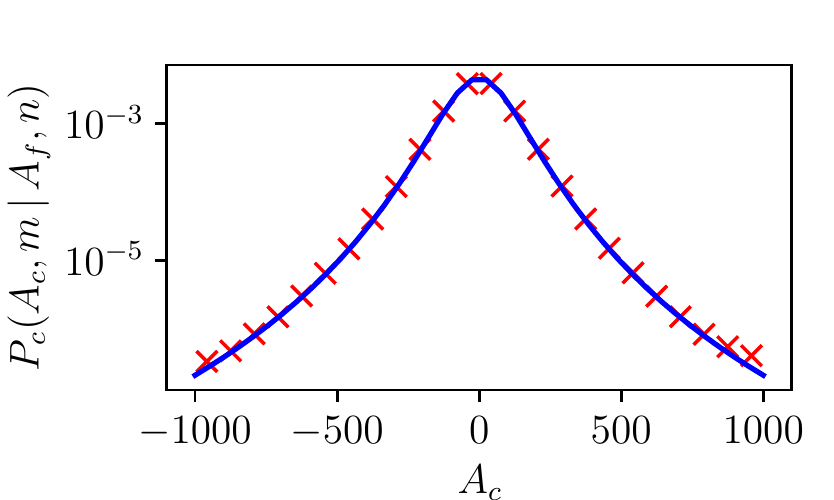}%
}\hfill
\caption{\textbf{Left panel:} A typical trajectory $x_{c,m}$ vs $m$ (top) and $A_{c,m}$ vs $m$ (bottom) of a Cauchy bridge random walk of $n=50$ steps with a zero area constraint $A_f=0$ generated by the effective jump distribution (\ref{eq:feffC}) with $\gamma=1$. \textbf{Right panel:} Position (top) and area (bottom) distributions at $m=25$ for a bridge random walk of $n=50$ steps with a zero area constraint. The marginal distributions, obtained numerically by sampling the trajectories from the effective jump distribution (\ref{eq:feffC}), are compared with the theoretical predictions for marginal distributions obtained from (\ref{marg_xc}), (\ref{marg_Ac}) and (\ref{eq:PcC}). The numerically obtained correlator using (\ref{eq:feffC}) $\langle(x_c-\mu_x)(A_c-\mu_A) \rangle \approx 37.809$ agrees well with the theoretical estimation $\approx 37.607$ obtained from (\ref{correl_xA_Cauchy}).}\label{fig:CRWconstrained}
\end{figure}

This effective jump distribution can then be used to generate Cauchy bridges with a fixed area. The distribution (\ref{eq:feffC}) is an example where the distribution cannot be sampled directly and where the ARS method comes in handy. To use the ARS method described in (\ref{eq:constant}), we need to find a constant $c_{m,n}(x_c,A_c,A_f) \geq 1$ that is independent of $\eta$. Substituting (\ref{eq:feffC}) in the inequality (\ref{eq:constant}), one finds that 
\bea \label{ineq_cmn} 
c_{m,n}(x_c,A_c,A_f) \geq \frac{Q(x_c+\eta,A_c+x_c+\eta,n-m-1)}{Q(x_c,A_c,n-m)} \;,
\eea
for all $\eta$. To find this $\eta$-independent constant $c_{m,n}$, we can replace the right hand side of this inequality (\ref{ineq_cmn}) by its maximal value. 
Using the integral representation of $Q(x,A,n-m-1)$ in the second line of (\ref{eq:PQcau}), the real part of its right hand side (note that its imaginary part vanishes) and using that $\cos(z) \leq 1$, one finds that $Q(x,A,n-m-1)\leq Q(0,A_f,n-m-1)$. Therefore one can choose the constant to be the maximal value of the right hand side of the inequality in (\ref{ineq_cmn}), which also happens to be greater than one, and is given by
\begin{align}
 c_{m,n}(x_c,A_c,A_f) &= \frac{Q(0,A_f,n-m-1)}{Q(x_c,A_c,n-m-1)} \\
 &= \frac{\int_{0}^{2\pi}d\theta\, \frac{1}{[\gamma\sum_{l=0}^{n-m-2} \, |\sin(\theta)-l\,\cos(\theta)|]^2}}{\int_{0}^{2\pi}d\theta\, \frac{1}{[ix_c\sin(\theta)+i(A_c-A_f)\cos(\theta)+ \gamma\sum_{l=0}^{n-m-1} \, |\sin(\theta)-l\,\cos(\theta)|]^2}}\,.\label{eq:cmnC}
\end{align}
Using (\ref{eq:paccept}), this yields the following acceptance probability
\begin{equation}
  p_{\text{accept}}(\eta,x_c,A_c,A_f,m,n) = 
  \frac{\int_{0}^{2\pi} \frac{d\theta}{[i(x_c+\eta)\sin(\theta)+i(A_c-A_f+x_c+\eta)\cos(\theta)+ \gamma\sum_{l=0}^{n-m-2} \, |\sin(\theta)-l\,\cos(\theta)|]^2}}{\int_{0}^{2\pi}\, \frac{d\theta}{[\gamma\sum_{l=0}^{n-m-2} \, |\sin(\theta)-l\,\cos(\theta)|]^2}}\,,\label{eq:pacceptC}
\end{equation}
which can be evaluated numerically. The ARS method can be used to generate the constrained trajectories (see left panel in figure \ref{fig:CRWconstrained}).
In the right panel in figure \ref{fig:CRWconstrained}, we computed numerically the marginal probability distributions of the position and the area at some intermediate time $m=25$, by generating trajectories from (\ref{eq:feffC}). This is compared to the theoretical marginal distributions of the position and the area which can be easily computed by substituting the free propagators (\ref{eq:PCf}) and (\ref{eq:PCb}) in (\ref{eq:Pcd}), which gives
\begin{align}
 &P_c(x_c,A_c,m\,|\,A_f,n) = \frac{1}{4\pi^2}\frac{1}{\int_{0}^{2\pi}d\theta\, \frac{1}{[\gamma\sum_{l=1}^{n} \, |\sin(\theta)+l\,\cos(\theta)|]^2}}\nonumber \\
&\times \int_{0}^{2\pi}d\theta\, \frac{1}{[ix_c\sin(\theta)+iA_c\cos(\theta)+ \gamma\sum_{l=1}^{m} \, |\sin(\theta)+l\,\cos(\theta)|]^2} \nonumber\\
&\times \int_{0}^{2\pi}d\theta\, \frac{1}{[ix_c\sin(\theta)+i(A_c-A_f)\cos(\theta)+ \gamma\sum_{l=0}^{n-m-1} \, |\sin(\theta)-l\,\cos(\theta)|]^2}\,,\label{eq:PcC}
\end{align}
which can be evaluated numerically straightforwardly. The marginal distributions $P_c(x_c,m\,|\,A_f,n)$ and $P_c(A_c,m\,|\,A_f,n)$
can then be obtained by integrating (\ref{eq:PcC}) over $x_c$ or $A_c$ respectively, as in (\ref{marg_xc}) and (\ref{marg_Ac}). In figure \ref{fig:CRWconstrained},
these marginal distributions, evaluated via numerical integration, are compared with the numerically sampled marginal distributions using the effective jump distribution in (\ref{eq:feffC}), showing a very good agreement.

As in the Gaussian
case, the two random variables $x_c$ and $A_c$ are correlated since the joint distribution does not factorise into the product of the two marginal distributions. Note that in the Gaussian case, the connected correlation matrix characterises the full bivariate distribution in (\ref{eq:PcG}). In contrast, in the Cauchy case in (\ref{eq:PcC}), it is not fully characterised by the correlation matrix. Nevertheless, there is a nonzero correlation between $x_c$ and $A_c$ in the Cauchy case, that can be computed as follows
\begin{align} \label{correl_xA_Cauchy}
\langle(x_c-\mu_x)(A_c-\mu_A) \rangle = \int_{-\infty}^\infty dx_c \, \int_{-\infty}^\infty dA_c \; (x_c-\mu_x) \, (A_c-\mu_A) \,P_c(x_c,A_c,m\,|\,A_f,n)\,.
\end{align}
where
\begin{subequations}
\begin{align}
  \mu_x &= \int_{-\infty}^\infty dx_c  \; x_c \,P_c(x_c,m\,|\,A_f,n)\,,\\
   \mu_A &= \int_{-\infty}^\infty dA_c \; A_c\,P_c(A_c,m\,|\,A_f,n)\,,
\end{align}
\label{eq:muxAC}
\end{subequations}
where $P_c(x_c,m\,|\,A_f,n)$ and $P_c(A_c,m\,|\,A_f,n)$
are the marginal distributions that can then be obtained by integrating (\ref{eq:PcC}) over $x_c$ or $A_c$ respectively, as in (\ref{marg_xc}) and (\ref{marg_Ac}). By substituting the joint distribution from (\ref{eq:PcC}), the double integral (\ref{correl_xA_Cauchy}) can be easily evaluated numerically and compared
to numerical simulations using the effective jump distribution (\ref{eq:feffC}), finding very good agreement.

\section{Generalisation to other global constraints}\label{sec:gen}

In the previous sections, we obtained a method to generate bridge trajectories with a fixed area. In this section, we generalise our construction to other global constraints. Going beyond the area, one could ask the more general question: How to generate Brownian paths $x(t)$ of duration $t_f$ with the value of a general observable $\mathcal{O}(t_f)=\int_0^{t_f} dt\,V[x(t)]$ fixed where $V(x)$ can be any arbitrary function? Such observables are usually referred to as functionals of Brownian motion (see for instance \cite{Majumdar05Ein,Perret15}). While it seems difficult to provide an exact answer for an arbitrary $V(x)$, there exist two specific examples, beyond the area where $V(x)=x$, for which we can make analytical progress: the occupation time of Brownian motion on the positive axis, which corresponds to $V(x)=\Theta(x)$, where $\Theta(x)$ is the Heaviside step function, and a ``generalised area'' which corresponds to $V(x)=x^n$, where $n$ is an integer. Below, we first show how to generate Brownian bridges with a fixed occupation time on the positive axis. We then outline the derivation for the generalised area and obtain explicit expression for the case of $n=2$. This latter case is of interest in the context of characterising the roughness of fluctuating $(1+1)$-dimensional interfaces \cite{Foltin94,Racz94}.

\subsection{Generating Brownian motion with a fixed occupation time on the positive axis}
\begin{figure}[t]
\subfloat{%
 \includegraphics[width=0.5\textwidth]{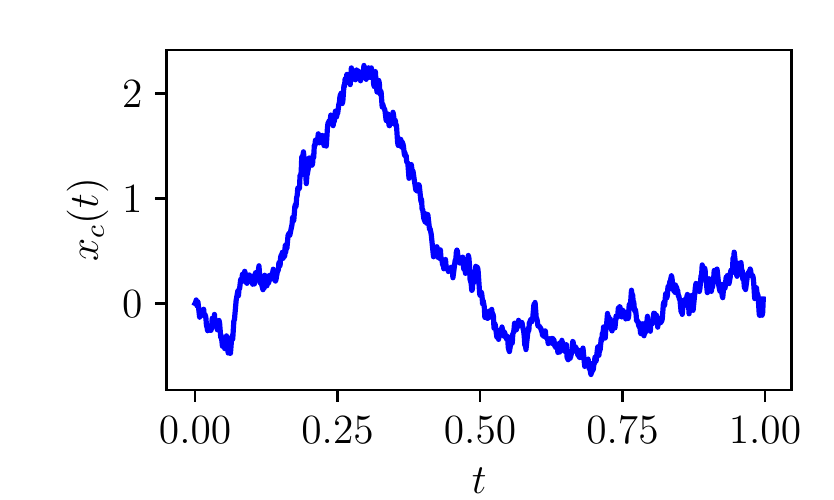}%
}\hfill
\subfloat{%
  \includegraphics[width=0.5\textwidth]{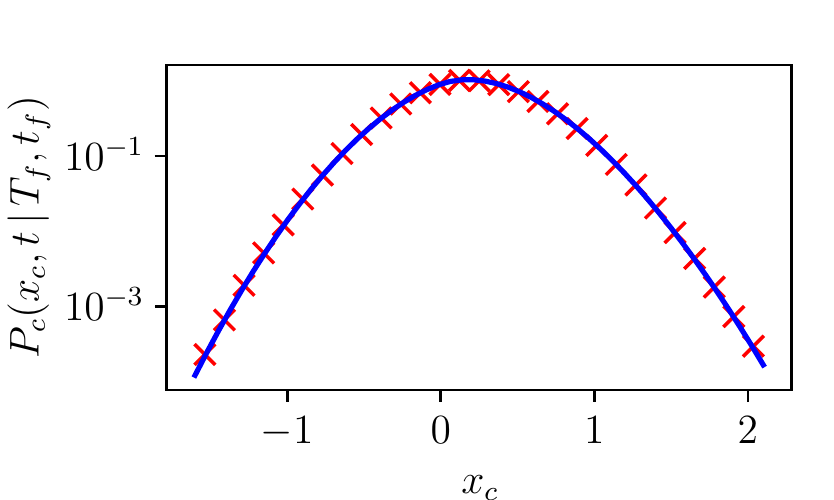}%
}\hfill\\
\subfloat{%
 \includegraphics[width=0.5\textwidth]{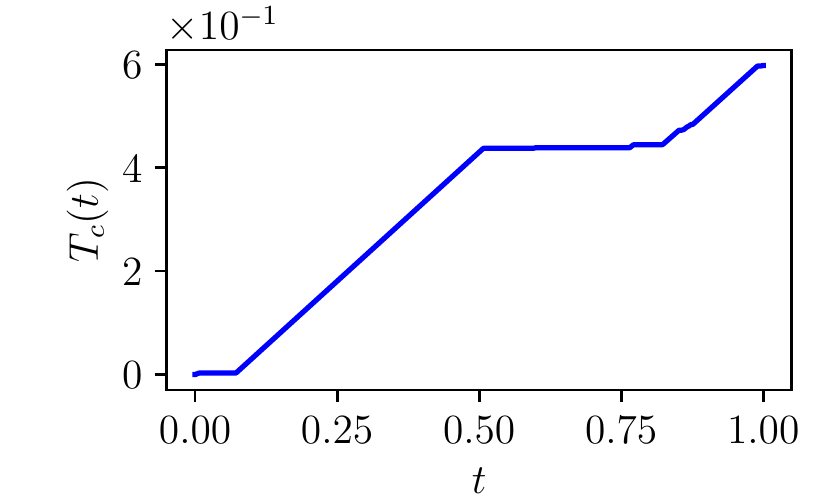}%
}\hfill
\subfloat{%
  \includegraphics[width=0.5\textwidth]{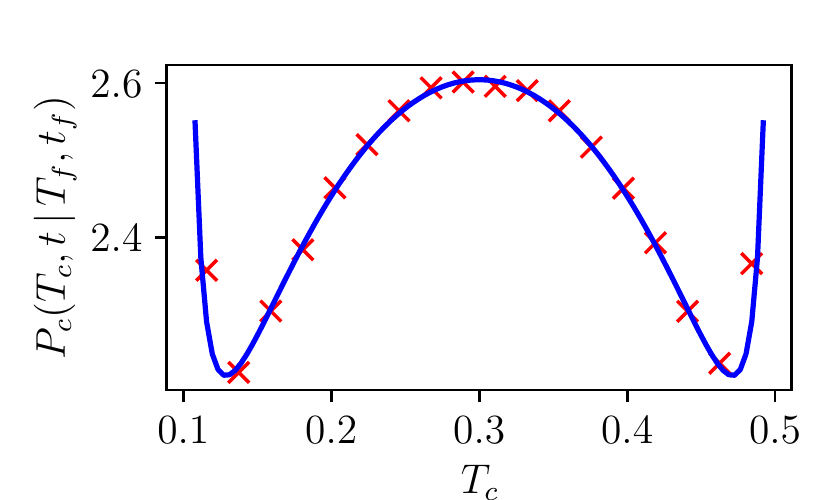}%
}\hfill
\caption{\textbf{Left panel:} A typical trajectory $x_c(t)$ vs $t$ (top) and $T_c(t)$ vs $t$ (bottom) of a bridge Brownian motion of duration $t_f=1$ with a fixed occupation time of $T_f=0.6$ generated by the effective Langevin equation (\ref{eq:effLBMt}) for $D=1$. \textbf{Right panel:} Marginal position (top) and occupation time (bottom) distributions at $t=t_f/2$ for a bridge Brownian motion of duration $t_f=1$ with a fixed occupation time of $T_f=0.6$. The distributions, obtained numerically by sampling the trajectories from the effective Langevin equation (\ref{eq:effLBMt}), are compared with the theoretical predictions in (\ref{eq:Pcbzt}).  The numerically obtained correlator using (\ref{eq:effLBMt}) is  $\langle( x_c-\mu_x)(T_c-\mu_T)\rangle \approx 2\times 10^{-4}$ and the theoretical one, obtained using (\ref{correl_xT}), is~$0$.}\label{fig:constrainedo}
\end{figure}

\begin{figure}[t]
    \begin{center}
      \includegraphics[width=0.6\textwidth]{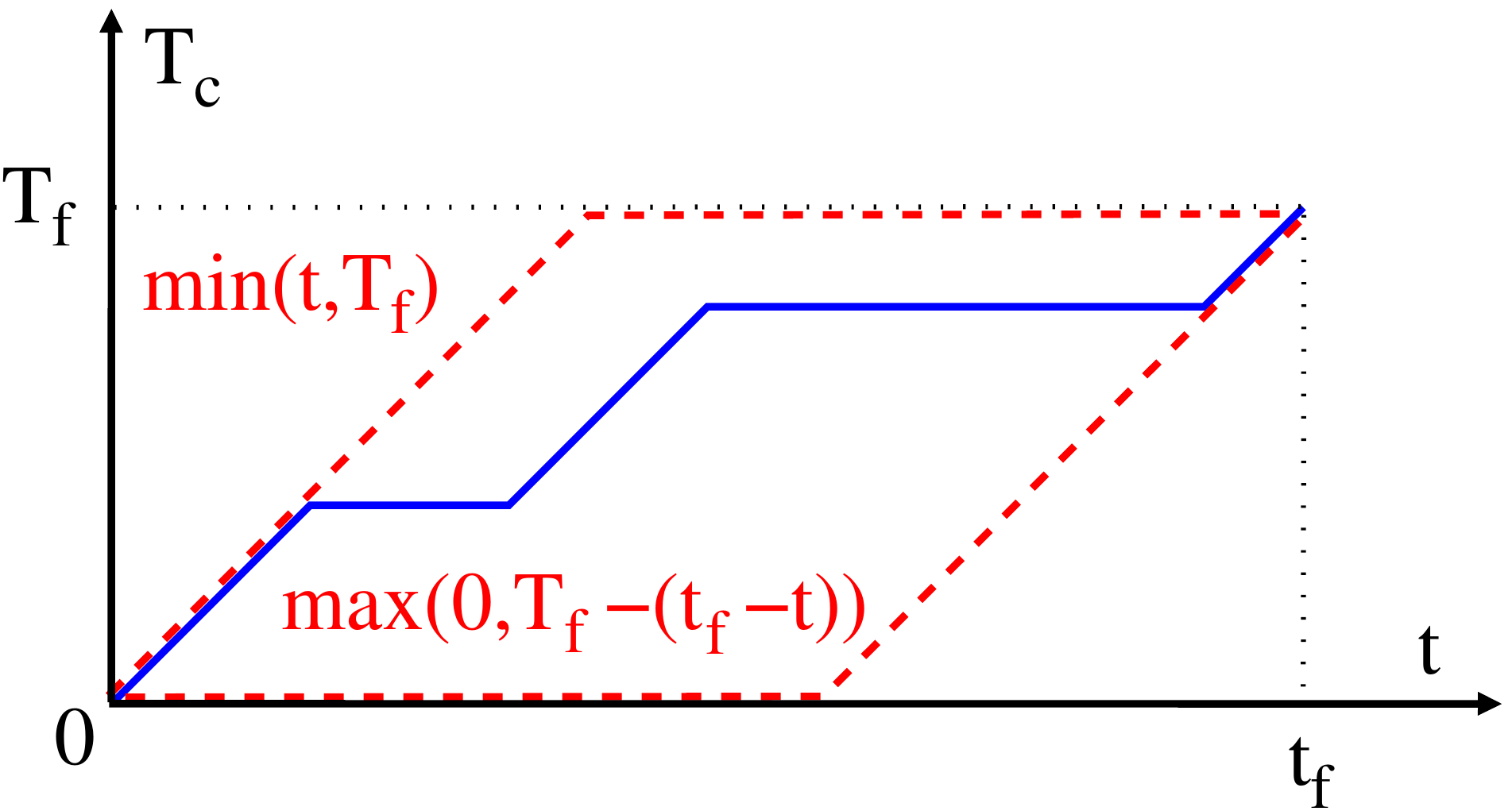}
      \caption{The constrained occupation time $T_c(t)$ as a function of time (blue line), which is a positive increasing function of time $t$ with a slope $0$ or $1$ and which must start at the origin and finish at $T_f$ at the time $t_f$, is constrained to remain between the top and bottom dashed red lines, respectively given by $\text{min}(t,T_f)$ and $\text{max}(0,T_f-(t_f-t))$.}
      \label{fig:Tc}
    \end{center}
  \end{figure}

The occupation time on the positive axis $T(t)=\int_0^{t} dt'\,\Theta[x(t')]$ of a Brownian path $x(t')$ corresponds to the total amount of time it has spent on the positive axis. For a free Brownian motion, the distribution of this time follows the well-known ``L\'evy's arcsine law'' \cite{Levy39} -- for generalisations to other stochastic processes see \cite{Lamperti58}. In physics, this observable is important in the context of stationary processes \cite{MB2002}, coarsening dynamics \cite{Dornic98,Newman98}, anomalous diffusive processes \cite{BBDG99,DeSmedt01,Dhar99}, blinking quantum dots \cite{Dahan03,MB05,barkai09} and spin glasses or disordered systems \cite{Majumdar02,MC02,SMC06,BB2007}. Below, we show how to generate exactly Brownian bridges $x_c(t)$ of duration $t_f$ with a fixed occupation time $T_c(t_f)=T_f$. As in the previous sections, it is convenient to consider the joint process
\begin{subequations}
\begin{align}
  \dot x_c(t) &= \sqrt{2D}\,\eta(t)\,,\\
  \dot T_c(t) &= \Theta(x_c(t))\,,
\end{align}
\label{eq:xtn}
\end{subequations}
\hspace*{-0.1cm}where $\Theta(x)$ is the Heaviside step function, i.e. $\Theta(x)=1$ if $x>0$ and $\Theta(x)=0$ otherwise. The constraints on the trajectories read
\begin{align}
  x(0)=x(t_f)=0\,,\quad T_c(t_f) = T_f\,.\label{eq:constt}
\end{align}
The constrained propagator $P_c(x_c,T_c,t\,|\,T_f,t_f)$, which is the probability distribution that the particle is located at $x_c$ with an occupation time $T_c$ at a time $t<t_f$ is given by the normalised product
\begin{align}
  P_c(x_c,T_c,t\,|\,T_f,t_f) = \frac{P(x_c,T_c,t)\,Q(x_c,T_c,t_f-t)}{P(0,T_f,t_f)}\,,\label{eq:Pct}
\end{align}
As in the case of the area observable discussed in Section \ref{sec:BM}, one can derive evolution equations for the two propagators 
 $P(x,T,t)$ and $Q(x,T,t)$. They read
\begin{subequations}
\begin{align}
\partial_t P(x, T,t) = D \partial^2_{x} P(x, T,t) -\Theta(x)\, \partial_ T P(x, T,t)\,,\label{eq:Peqxit}\\
\partial_t Q(x, T,t) = D \partial^2_{x} Q(x, T,t) +\Theta(x)\, \partial_ T Q(x, T,t)\;.\label{eq:Qeqxit}
\end{align}
\end{subequations}
One can then generate constrained trajectories using the effective Langevin equation
\begin{subequations}
\begin{align}
  \dot x_c(t) &= \sqrt{2\,D}\,\eta(t) + 2\,D\,\partial_x\ln[Q(x_c(t),T_c(t),t_f-t)]\,,\\
  \dot T_c(t) &= \Theta[x_c(t)]\,,
\end{align}
\label{eq:effgT}
\end{subequations}
where $Q(x, T,t)$ is the solution of (\ref{eq:Qeqxit}) with the initial condition $Q(x, T,0)=\delta(x)\delta( T-T_f)$. The solution of this equation is given by (see  \ref{app:time})
\begin{align}
  Q(x, T,t) = \frac{1}{2\pi\sqrt{Dt^3}}\,\mathcal{F}\left(y=\frac{|x|}{\sqrt{Dt}},\nu = \left(\frac{T_f- T}{t-(T_f- T)}\right)^{\text{sign}(x)}\right)\,,\label{eq:Qdimtm}
\end{align}
where
\begin{align}
  \mathcal{F}(y,\nu)  &=  e^{-\frac{y^2}{4}}\left[\sqrt{\nu}\,y e^{-\frac{y^2}{4\nu}}-\frac{\sqrt{\pi}}{2}(y^2-2)\,\text{erfc}\left(\frac{y}{2\sqrt{\nu}} \right)\right]\,, \label{eq:Fdimm}
\end{align}
where $\text{erfc}(z)=1-\frac{2}{\sqrt{\pi}}\int_0^z e^{-t^2}dt$ is the complementary error function. Inserting this expression into the effective Langevin equation (\ref{eq:effgT}), we find
\begin{subequations}
\begin{align}
  \dot x_c(t) &= \sqrt{2\,D}\,\eta(t)+\,\frac{2\,\text{sign}(x_c(t))\sqrt{D}}{\sqrt{t_f-t}} \nonumber\\
  & \times \partial_y \ln\left[\mathcal{F}\left(y=\frac{|x_c(t)|}{\sqrt{D(t_f-t)}},\nu = \left(\frac{T_f-T_c(t)}{(t_f-t)-(T_f-T_c(t))}\right)^{\text{sign}(x_c(t))}\right)\right] \,,\\
  \dot T_c(t) &= \Theta[x_c(t)]\,,
\end{align}
\label{eq:effLBMt}
\end{subequations}
where the derivative $\partial_y \ln(\mathcal{F}(y,\nu))$ is given by
\begin{align}
 \partial_y \ln(\mathcal{F}(y,\nu)) = \frac{\sqrt{\pi \nu } y \left(y^2-6\right) e^{\frac{y^2}{4 \nu }} \text{erfc}\left(\frac{y}{2
   \sqrt{\nu }}\right)-2 \nu  \left(y^2-2\right)-4}{4 \nu  y-2 \sqrt{\pi\nu } \left(y^2-2\right)
   e^{\frac{y^2}{4 \nu }} \text{erfc}\left(\frac{y}{2 \sqrt{\nu }}\right)}\,.\label{eq:Fdimdm}
\end{align}
This effective Langevin equation can be used to generate constrained trajectories (see figure \ref{fig:constrainedo}).
 In the right panel in figure \ref{fig:constrainedo}, we computed numerically the marginal probability distributions of the position and the occupation time at some intermediate time $t=t_f/2$, by generating trajectories from (\ref{eq:effLBMt}). This is compared to the theoretical marginal distributions of the position and the occupation time for the constrained process which can be easily computed by substituting the free propagators $Q(x_c,T_c,t)$ and $P(x_c,T_c,t)=Q(x_c,T_f-T_c,t)$ from (\ref{eq:Qdimtm}) in (\ref{eq:Pct}), which gives
\begin{align}
 P_c(x_c,T_c,t\,|\,T_f,t_f) &=  \frac{t_f^{3/2}}{4\sqrt{D\pi^3t^3(t_f-t)^3}}\,\mathcal{F}\left(y=\frac{|x_c|}{\sqrt{Dt}},\nu = \left(\frac{T_c}{t-T_c}\right)^{\text{sign}(x_c)}\right)\nonumber\\
  &\times\mathcal{F}\left(y=\frac{|x_c|}{\sqrt{D(t_f-t)}},\nu = \left(\frac{T_f-T_c}{(t_f-t)-(T_f-T_c)}\right)^{\text{sign}(x_c)}\right)\,,\label{eq:Pcbzt}
\end{align}
where we used that the numerator in (\ref{eq:Pct}) is $P(0,T_f,t_f)=\frac{\sqrt{\pi}}{2\sqrt{Dt_f^3}}$.
 The marginal distributions can then be obtained by integrating (\ref{eq:Pcbzt}) over $x_c$ or $T_c$ respectively
 \begin{subequations}
 \begin{align}
   P_c(x_c,t\,|\,T_f,t_f) &= \int_{\text{max}(0,T_f-(t_f-t))}^{\text{min}(t,T_f)} P_c(x_c,T_c,t\,|\,T_f,t_f)\,dT_c \,,\label{eq:margpcxta} \\
   P_c(T_c,t\,|\,T_f,t_f) &= \int_{-\infty}^\infty P_c(x_c,T_c,t\,|\,T_f,t_f)\,dx_c \,, \label{eq:margpcxtb}
 \end{align}
 \label{eq:margpcxt}
 \end{subequations}
\hspace*{-0.1cm}where the integral over $T_c$ spans from $\text{max}(0,T_f-(t_f-t))$ to $\text{min}(t,T_f)$  (see also figure \ref{fig:Tc}). These limits can be understood as follows. We note that $T_c = \int_0^t \Theta(x(t'))\, dt'$, which is just the occupation time till $t$. Let us define $\overline{T_c} = \int_{t}^{t_f} \Theta(x(t'))\, dt'$. Clearly $T_c + \overline{T_c} = T_f$, by definition. We note further that $\overline{T_c}$ clearly satisfies the inequality $\overline{T_c} \leq t_f-t$. Therefore, this implies that $T_c \geq T_f - (t_f-t)$ which, combined with the fact that $T_c \geq 0$, leads
to the lower bound $T_c \geq \max(0, T_f - (t_f-t))$ in (\ref{eq:margpcxta}). The upper bound is easier since, by definition, $T_c  \leq t$ and also $T_c \leq T_f$ since $T_c = \int_0^t \Theta(x(t'))\, dt'$ is an increasing function of $t$. Therefore $T_c \leq \min(t,T_f)$. 
These marginal distributions in (\ref{eq:margpcxta}) and (\ref{eq:margpcxtb}) are plotted in the right panel in figure \ref{fig:constrainedo} and compared to numerical simulations using the 
effective evolution equations (\ref{eq:effLBMt}), finding excellent agreement. As an additional check, we compare the numerical correlation $\langle (x_c -\mu_x)(T_c - \mu_T) \rangle$ and the theoretical one to probe the joint distribution of $x_c$ and $T_c$ beyond its marginal distributions. The connected correlation between $x_c$ and $T_c$ can be computed as follows
\begin{align} \label{correl_xT}
\langle(x_c-\mu_x)(T_c-\mu_T) \rangle = \int_{-\infty}^\infty dx_c \, \int_{\text{max}(0,T_f-(t_f-t))}^{\text{min}(t,T_f)} dT_c \; (x_c-\mu_x) \, (T_c-\mu_T) \,P_c(x_c,T_c,t\,|\,T_f,t_f)\,.
\end{align}
where
\begin{subequations}
\begin{align}
  \mu_x &= \int_{-\infty}^\infty  \; x_c \,P_c(x_c,t\,|\,T_f,t_f)\, dx_c\,,\\
   \mu_T &= \int_{\text{max}(0,T_f-(t_f-t))}^{\text{min}(t,T_f)}\; T_c\,P_c(T_c,t\,|\,T_f,t_f)\,dT_c\,,
\end{align}
\label{eq:muxTB}
\end{subequations}
\hspace*{-0.1cm}where $P_c(x_c,t\,|\,T_f,t_f)$ and $P_c(A_c,t\,|\,T_f,t_f)$
are the marginal distributions given in (\ref{eq:margpcxt}). By substituting the joint distribution from (\ref{eq:Pcbzt}), the double integral (\ref{correl_xT}) can be easily evaluated numerically and compared
to numerical simulations using the effective Langevin equation (\ref{eq:effLBMt}), finding very good agreement.

\subsection{Generating Brownian motion with a fixed generalised area}
\begin{figure}[t]
\subfloat{%
 \includegraphics[width=0.5\textwidth]{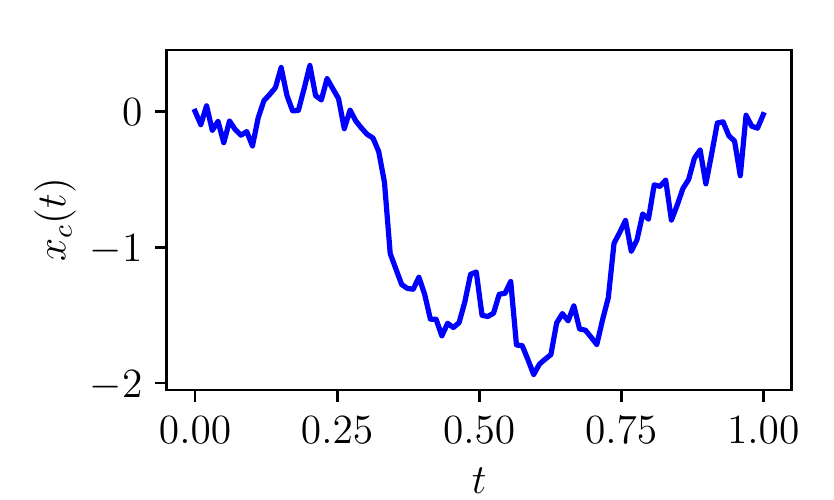}%
}\hfill
\subfloat{%
 \includegraphics[width=0.5\textwidth]{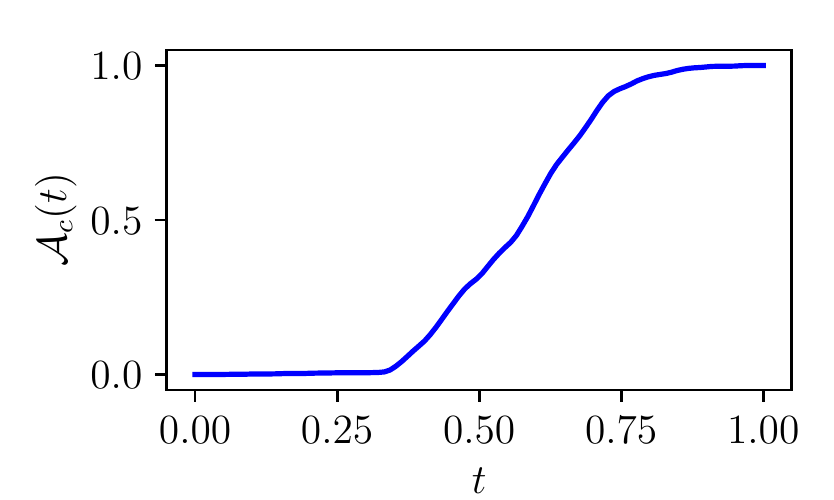}%
}\hfill
\caption{\textbf{Left panel:} A typical trajectory $x_c(t)$ vs $t$ (left) and $\mathcal{A}_c(t)$ vs $t$ (right) of a bridge Brownian motion of duration $t_f=1$ with a fixed quadratic area $\mathcal{A}_{f}= \int_0^{t_f} dt\, x_c^2(t) = 1$ generated by the effective Langevin equation (\ref{eq:effLBMg}) with $D=1$. }\label{fig:constrainedXi}
\end{figure}

The generalised area $\mathcal{A}(t) = \int_0^{t} dt'\, x^n(t')$ of a Brownian path $x(t')$ extends the notion of area discussed in the previous sections, to an arbitrary power of $n$. The case of $n=2$ plays an important role in the context of fluctuating $(1+1)$-dimensional interfaces as it characterises the roughness of the profile \cite{Foltin94,Racz94}. The distribution of this random variable $\mathcal{A}(t)$ has also been studied extensively when $x(t')$ represents the Ornstein-Uhlenbeck process \cite{MB2002, Meer2010, NT2018,NRSmith}. Here, we only focus only on a Brownian bridge of total duration $t_f$ and subject to the global constraint that the generalized area under the bridge has a fixed value $\mathcal{A}_f =  \int_0^{t_f} dt\, x^n(t)$.  
We denote this constrained process by $x_c(t)$ and outline below the derivation of the effective Langevin equation. As in the previous sections, it is convenient to consider the joint process
\begin{subequations}
\begin{align}
  \dot x_c(t) &= \sqrt{2D}\,\eta(t)\,,\\
  \dot{\mathcal{A}_{c}}(t) &= x^n(t)\,,
\end{align}
\label{eq:xAn}
\end{subequations}
with the following bridge and generalised area constraints
\begin{align}
  x_c(0)=x_c(t_f)=0\,,\quad \mathcal{A}_{c}(t_f) = \mathcal{A}_{f}\,.\label{eq:constg}
\end{align}
The constrained propagator $P_c(x_c,\mathcal{A}_{c},t\,|\,t_f,A_f)$, which is the probability distribution that the particle is located at $x_c$ with a generalised area $\mathcal{A}_{c}$ at a time $t<t_f$ is given by the normalised product
\begin{align}
  P_{c}(x_{c},\mathcal{A}_{c},t\,|\,\mathcal{A}_{f},t_f) = \frac{P(x_{c},\mathcal{A}_{c},t)\,Q(x_{c},\mathcal{A}_{c},t_f-t)}{P(0,\mathcal{A}_{c},t_f)}\;. \label{eq:Pc2}
\end{align}
 As in the previous sections, one can derive evolution equations for the two propagators $P(x,\mathcal{A},t)$ and $Q(x,\mathcal{A},t)$. They read
\begin{subequations}
\begin{align}
\partial_t P(x,\mathcal{A},t) = D \partial_x^2 P(x,\mathcal{A},t) -x^n \partial_\mathcal{A} P(x,\mathcal{A},t)\,,\label{eq:Peqxi}\\
\partial_t Q(x,\mathcal{A},t) = D \partial_x^2 Q(x,\mathcal{A},t) +x^n \partial_\mathcal{A} Q(x,\mathcal{A},t)\,. \label{eq:Qeqxi}
\end{align}
\end{subequations}
One can then generate constrained trajectories using the effective Langevin equation
\begin{subequations}
\begin{align}
  \dot x_c(t) &= \sqrt{2\,D}\,\eta(t) + 2\,D\,\partial_x\ln[Q(x_c(t),\mathcal{A}_{c}(t),t_f-t)]\,,\label{eq:effgLa} \\
  \dot{\mathcal{A}_{c}}(t) &= x^n_c(t)\,, \label{eq:effgLb}
\end{align}
\label{eq:effgL}
\end{subequations}
\hspace*{-0.1cm}where $Q(x,\mathcal{A},t)$ is the solution of the equation (\ref{eq:Qeqxi}) with the initial condition $Q(x,\mathcal{A},0)=\delta(x)\delta(\mathcal{A}-\mathcal{A}_f)$. Note that for $n=1$, we can compute $Q(x,\mathcal{A},t)$ explicitly, as was shown in detail in Section \ref{sec:BM}. One then verifies that, for $n=1$, (\ref{eq:effgLa}) and (\ref{eq:effgLb}) indeed reduce to (\ref{eq:effLBMa}) and (\ref{eq:effLBMb}). For arbitrary $n > 1$, it is difficult to solve explicitly these equations (\ref{eq:Peqxi}) and (\ref{eq:Qeqxi}). However, for $n=2$, it is still possible to make analytical progress (see \ref{app:quad}). In this case, the solution of (\ref{eq:Qeqxi}) reads 
\begin{align}
   Q(x,\mathcal{A},t) = \frac{1}{\sqrt{2\pi D^3\,t^5}}\, \mathcal{G}\left(y=\frac{x}{\sqrt{Dt}},z=\frac{\mathcal{A}_f-\mathcal{A}}{D t^2}\right)\,,\label{eq:Qdim}
\end{align}
with the scaling function 
 \begin{align}
  \mathcal{G}(y,z) = \int_\Gamma \frac{du}{2\pi i} \, u^{1/4}\,\frac{e^{z\,u-\frac{y^2}{2}\sqrt{u}\,\coth(2\sqrt{u})}}{\sqrt{\sinh(2\sqrt{u})}}\,,\label{eq:Gdim2}
\end{align} 
where $\Gamma$ is a Bromwich contour in the complex $u$-plane. 
Inserting this expression into the effective Langevin equations (\ref{eq:effgL}), we find
\begin{subequations}
\begin{align}
  \dot x_c(t) &= \sqrt{2\,D}\,\eta(t) - \frac{2x_c(t)}{(t_f-t)} \frac{ \int_\Gamma du \, \frac{u^{3/4}\,\coth(2\sqrt{u})}{\sqrt{\sinh(2\sqrt{u})}}\,\exp\left(\frac{\mathcal{A}_{f}-\mathcal{A}_{c}(t)}{D (t_f-t)^2}\,u-\frac{x_c^2(t)\sqrt{u}\,\coth(2\sqrt{u})}{2D(t_f-t)}\right)}{ \int_\Gamma du \, \frac{u^{1/4}}{\sqrt{\sinh(2\sqrt{u})}}\exp\left(\frac{\mathcal{A}_{f}-\mathcal{A}_{c}(t)}{D (t_f-t)^2}\,u-\frac{x_c^2(t)\sqrt{u}\,\coth(2\sqrt{u})}{2D(t_f-t)}\right)}\,,\label{eq:effLBMga} \\
  \dot{\mathcal{A}_{c}}(t) &= x^2_c(t)\,, \label{eq:effLBMgb}
\end{align}
\label{eq:effLBMg}
\end{subequations}
This effective Langevin equation can be used to generate constrained trajectories with moderate accuracy (see figure \ref{fig:constrainedXi}). This is because the integrals in the complex $u$-plane in (\ref{eq:effLBMga}) are hard to evaluate numerically with high accuracy. 

\section{Summary and outlook}\label{sec:ccl}
In this work, we studied continuous-time and discrete-time bridge random walks in the presence of a time-integrated constraint. In the case of continuous-time Brownian motion, we developed an effective Langevin equation to generate Brownian bridges with a fixed area under their trajectory. In the case of discrete-time random walks, we obtained an effective jump distribution that implicitly accounts for the constraint. To illustrate our method, we provided examples with random walks having a Gaussian or Cauchy jump distribution. We further generalised our approach to the case of Brownian motion by studying other time-integrated constraints such as a fixed occupation time on the positive axis and a fixed generalised area. 
It would be interesting to investigate if the present method can be extended to higher dimensions, non-Markovian processes and to multi-particle processes with interactions. 

Lately, there has been increasing interest in generating rare events with a not so small probability. They appear in many problems in out-of-equilibirum 
systems where such rare events are characterised by large deviations probabilities, which are typically extremely tiny. There has been recent progress
in developing efficient numerical algorithms, using importance sampling, that allow to measure such rare events with probabilities as small as $10^{-100}$ \cite{Hart2015, HLMRS18}! 
In the context of systems out-of-equilibrium, new algorithms, inspired by reinforcement learning and machine learning approaches, have been developed \cite{Gar2018, Rose21, Rose21area}. Here, we have
provided a simple single-particle stochastic process in the presence of time-integrated dynamical constraint and we have shown that it can be generated	very simply
using an effective but exact Langevin equation. It would be interesting to see what this result may imply for algorithms based on reinforcement learning where this question
was recently raised \cite{Rose21}.

\section*{Acknowledgments}
This work was partially supported by the Luxembourg National Research Fund (FNR) (App. ID 14548297).

\appendix 

\section{Free propagator of the joint position and area for Brownian motion}\label{app:area}
In this appendix, we derive the free propagators $P(x,A,t)$ and $Q(x,A,t)$ of the joint position $x$ and the area $A$ at time $t$ for Brownian motion. We start with $P(x,A,t)$ and we solve the equation (\ref{eq:Peq}), namely
\begin{align}
\partial_t P(x,A,t) = D \partial_x^2 P(x,A,t) -x\, \partial_A P(x,A,t)\,,\label{eq:Peq2}
\end{align}
with the initial condition $P(x,A,0)=\delta(x)\delta(A)$. Given the equation of motion (\ref{eq:eom})-(\ref{eq:Ac}), we know that $P(x,A,t)$ is a bi-variate Gaussian distribution, i.e.,
\begin{align}
  P(x,A,t)=\frac{1}{ \mathcal{N}\,D t^2}\,\exp\left( \frac{a_{1}A^2}{Dt^3}+  \frac{a_{2}x^2}{Dt} + \,\frac{a_{3}A x}{Dt^2}\right)\,,\label{eq:Pxaans}
\end{align}
where the normalisation factor $\mathcal{N}$ and the constants $a_1$, $a_2$ and $a_3$ have to be determined and the factors $D$ and $t$ have been set by dimensional analysis. Inserting this ansatz into the differential equation (\ref{eq:Peq2}), one finds
\begin{align}
\frac{A^2}{Dt^3} \left(3 a_1+a_3^2\right)+2 \frac{A x}{Dt^2} (-a_1+2 a_2 a_3+a_3)+\frac{x^2}{Dt} \left(4
   a_2^2+a_2-a_3\right)+2 (a_2+1)=0\,.\label{eq:conda123}
\end{align}
This equation must be true for all $A$ and $x$ which implies the following set of equations 
\begin{subequations}
\begin{align}
  3 a_1+a_3^2 &= 0\,,\\
  -a_1+2 a_2 a_3+a_3 &=0\,,\\
  4a_2^2+a_2-a_3 &= 0\,,\\
  a_2+1 &= 0\,.
\end{align}
\label{eq:asys}
\end{subequations}
The solution is
\begin{align}
  a_1 = -3\,,\quad a_2=-1\,,\quad a_3=3\,.\label{eq:assol}
\end{align}
Plugging these values into the distribution (\ref{eq:Pxaans}) and using the normalisation condition $\int_{-\infty}^\infty dx\int_{-\infty}^\infty dA P(x,A,t)=1$, we find $\mathcal{N}=2\pi/\sqrt{3}$ so that the solution is
\begin{align}
  P(x,A,t)=\frac{\sqrt{3}}{ 2\pi\,D t^2}\,\exp\left(-\frac{1}{D}\left[ \frac{3A^2}{t^3}+  \frac{x^2}{t} -\frac{3A x}{t^2}\right]\right)\,. \label{eq:PsolxA}
\end{align}
The propagator $Q(x,A,t)$ satisfying (\ref{eq:back}) together with the initial condition $Q(x,A,t=0)=\delta(A-A_f)\delta(x)$ is simply given by
\begin{align}
  Q(x,A,t) = P(x,A_f-A,t)\,,\label{eq:timrevqA}
\end{align}
where we used the time reversibility of Brownian motion. This yields the expression given in (\ref{eq:Qs}) in the text.

\section{Free propagator of the joint position and area for a Gaussian random walk}
\label{app:xAG}
In this appendix, we derive the free propagators $P(x,A,m)$ and $Q(x,A,m)$ of the joint position $x$ and area $A$ at step $m$ given in (\ref{eq:Qf}) for the case of a Gaussian random walk. The Fourier transform of the Gaussian jump distribution (\ref{eq:fG}) is given by
\begin{align}
  \hat f(k) = \int_{-\infty}^\infty d\eta\,e^{i\eta k}\frac{1}{\sqrt{2\pi\sigma^2}}e^{-\frac{\eta^2}{2\sigma^2}} = e^{-\sigma^2 k^2}\,.\label{eq:fourierG}
\end{align}
Inserting this expression in the propagators (\ref{eq:Qf}) gives
\begin{subequations}
\begin{align}
P(x,A,m) &= \int_{-\infty}^{\infty}\int_{-\infty}^{\infty} \frac{dk}{2\pi} \,\frac{d\lambda}{2\pi} \,e^{-ikx-i\lambda A} \exp\left(-\sigma^2\sum_{l=1}^{m} \,(k+l\,\lambda)^2\right)\,,\\
  Q(x,A,m) &= \int_{-\infty}^{\infty}\int_{-\infty}^{\infty} \frac{dk}{2\pi} \,\frac{d\lambda}{2\pi} \,e^{-ikx-i\lambda (A-A_f)} \exp\left(-\sigma^2\sum_{l=0}^{m-1} \,(k-l\,\lambda)^2\right)\,.
\end{align}
\label{eq:PQGRW1}
\end{subequations}
\hspace*{-0.1cm} By expanding the squares and performing the sums in the argument of the exponentials one obtains
\begin{subequations}
\begin{align}
 P(x,A,m) &= \int_{-\infty}^{\infty}\int_{-\infty}^{\infty} \frac{dk}{2\pi} \,\frac{d\lambda}{2\pi} \,e^{-ikx-i\lambda A} \exp\Bigg[-\sigma^2\Big(k^2 m+\lambda  k m (m+1)\nonumber\\
&\hspace{16em}+\frac{1}{6} \lambda ^2 m (m+1) (2 m+1)\Big)\Bigg]\,,\\
Q(x,A,m) &= \int_{-\infty}^{\infty}\int_{-\infty}^{\infty} \frac{dk}{2\pi} \,\frac{d\lambda}{2\pi} \,e^{-ikx-i\lambda (A-A_f)} \exp\Bigg[-\sigma^2\Big(k^2m- \lambda  km (m-1)\nonumber\\
&\hspace{16em}+\frac{1}{6}\lambda ^2m (m-1) (2 m-1)\Big)\Bigg]\,.
\end{align}
\label{eq:PQGRW2}
\end{subequations}
\hspace*{-0.1cm} Upon performing the Gaussian integrals, we recover the expressions (\ref{eq:PQGRW}) displayed in the main text.

\section{Free propagator of the joint position and occupation time on the positive axis for Brownian motion}\label{app:time}
In this appendix, we derive the free propagators $P(x, T,t)$ and $Q(x, T,t)$ of the joint position $x$ and occupation time $T<t$ on the positive axis at time $t$ for Brownian motion. We start with the propagator $P(x, T,t)$ whose equation is given in (\ref{eq:Peqxit}) and perform a double Laplace transform with respect to the variable $t$ and $ T$:
\begin{align}
  s \tilde P(x,\lambda,s) - \delta(x) = D \partial_x^2 \tilde P(x,\lambda,s) - \Theta(x)\lambda \tilde P(x,\lambda,s)\,,\label{eq:occtl}
\end{align}
where $\tilde P(x,\lambda,s)=\int_0^\infty d T \int_0^\infty dt P(x, T,t) e^{-st-\lambda  T} $ and where we used $P(x,0,t)=0$, because the Brownian motion will cross the origin immediately if it starts from the origin initially $P(x, T,0)=\delta(x)\delta( T)$. The differential equation (\ref{eq:occtl}) can be solved for $x>0$  and $x<0$ separately, where the Dirac delta term is absent. Using the fact that the solution must decay at $\pm \infty$, we find
\begin{align}
  \tilde P(x,\lambda,s) = \left\{\begin{array}{ll}
    A \,e^{x\sqrt{\frac{s}{D}}}\,,& x<0\,,\\
    B \,e^{-x\sqrt{\frac{s+\lambda}{D}}}\,,& x>0\,,
  \end{array}\right.\label{eq:gensol}
\end{align}
where $A$ and $B$ are integration constants that we fix by imposing the continuity of the solution at $x=0$, which gives $A=B$. Besides, by integrating  (\ref{eq:occtl}) around an infinitesimal interval centered around the origin one obtains a second relation, namely
\begin{align}
  -1 = D[\partial_x \tilde P(0^+,\lambda,s) - \partial_x \tilde P(0^-,\lambda,s)]\,,\label{eq:intdx}
\end{align}
from which we get 
\begin{align}
  A=B=\frac{1}{\sqrt{D}(\sqrt{s+\lambda}+\sqrt{s})}\,.\label{eq:ABint}
\end{align}
The double Laplace transform is therefore given by
\begin{align}
 \tilde P(x,\lambda,s)  = \frac{1}{\sqrt{D}(\sqrt{s+\lambda}+\sqrt{s})} \times \left\{\begin{array}{ll}
    e^{x\sqrt{\frac{s}{D}}}\,,& x<0\,,\\
    e^{-x\sqrt{\frac{s+\lambda}{D}}}\,,& x>0\,.
  \end{array}\right.\label{eq:ddsol}
\end{align}
We will now invert this double Laplace transform. Let us start with the case $x<0$. The propagator reads
\begin{align}
  P(x, T,t) =\int_\Gamma \frac{ds}{2\pi i}\,e^{st}\int_\Gamma \frac{d\lambda}{2\pi i} e^{\lambda  T}\frac{1}{\sqrt{D}(\sqrt{s+\lambda}+\sqrt{s})}e^{x\sqrt{\frac{s}{D}}}\,,\quad x<0\,,\label{eq:lapinv}
\end{align}
where $\Gamma$ is the usual Bromwich contour. The inversion with respect to $\lambda$ yields
\begin{align}
  P(x, T,t) =\int_\Gamma \frac{ds}{2\pi i}e^{st} \frac{e^{x\sqrt{\frac{s}{D}}}}{\sqrt{D}}\left(\frac{e^{-s T}}{\sqrt{\pi T}} - \sqrt{s}\,\text{erfc}(\sqrt{s T})\right)\,,\quad x<0\,,\label{eq:lapinv2}
\end{align}
where $\text{erfc}(z)=1-\frac{2}{\sqrt{\pi}}\int_0^z e^{-t^2}dt$ is the complementary error function. Given that the integrand in (\ref{eq:lapinv2}) is a product of two functions whose inverse Laplace transforms are given by
\begin{subequations}
\begin{align}
  \int_\Gamma \frac{ds}{2\pi i}e^{st} \,\frac{e^{x\sqrt{\frac{s}{D}}}}{\sqrt{D}} &= \frac{-x\,e^{-\frac{x^2}{4Dt}}}{\sqrt{4\pi Dt^3}}\,,\\
   \int_\Gamma \frac{ds}{2\pi i}e^{st} \left(\frac{e^{-s T}}{\sqrt{\pi T}} - \sqrt{s}\,\text{Erfc}(\sqrt{s T})\right) &= \frac{\Theta(t- T)}{2\sqrt{\pi}t^{3/2}}\,,
\end{align}
\label{eq:lapinve}
\end{subequations}
the inverse Laplace transform (\ref{eq:lapinv2}) will be a convolution of these two functions, i.e, 
\begin{align}
   P(x, T,t) &= \int_{0}^t dt' \, \frac{-x\,e^{-\frac{x^2}{4Dt'}}}{\sqrt{4\pi Dt'^3}} \times \frac{\Theta(t-t'- T)}{2\sqrt{\pi}(t-t')^{3/2}} \nonumber \\
   & = \int_{0}^{t- T} dt'\frac{-x\,e^{-\frac{x^2}{4Dt'}}}{4\pi\sqrt{ Dt'^3(t-t')^3}}\,,\quad x<0\,.\label{eq:Psolt}
\end{align}
By space reflexion symmetry of Brownian motion, the solution for $x>0$ will be given by $P(x, T,t) =P(-x,t- T,t)$. In terms of dimensionless variables variables, the solution can be conveniently written as
\begin{align}
  P(x, T,t) = \frac{1}{2\pi\sqrt{Dt^3}}\mathcal{F}\left(y=\frac{|x|}{\sqrt{Dt}},\nu = \left(\frac{ T}{t- T}\right)^{\text{sign}(x)}\right)\,,\label{eq:Pdimt}
\end{align}
where
\begin{align}
  \mathcal{F}(y,\nu) &= \frac{y}{2} \int_0^{1-\frac{1}{1+\nu}} du \frac{e^{-\frac{y^2}{4u}}}{\sqrt{u^3(1-u^3)}}\,,\\
  &= e^{-\frac{y^2}{4}}\left[\sqrt{\nu}\,y e^{-\frac{y^2}{4\nu}}-\frac{\sqrt{\pi}}{2}(y^2-2)\,\text{erfc}\left(\frac{y}{2\sqrt{\nu}} \right)\right]\,,\label{eq:Fdim}
\end{align}
where we further performed the integral by a change of variables $z=1\sqrt{u}$. As a check, one can verify that integrating over $x$ in (\ref{eq:Pdimt}) yields the ``Arcsine'' law, while integrating over $ T$ gives the usual Gaussian distribution. Note that the derivative $\partial_y \ln(\mathcal{F}(y,\nu))$ is given by
\begin{align}
  \partial_y \ln(\mathcal{F}(y,\nu)) = \frac{\sqrt{\pi \nu } y \left(y^2-6\right) e^{\frac{y^2}{4 \nu }} \text{erfc}\left(\frac{y}{2
   \sqrt{\nu }}\right)-2 \nu  \left(y^2-2\right)-4}{4 \nu  y-2 \sqrt{\pi\nu } \left(y^2-2\right)
   e^{\frac{y^2}{4 \nu }} \text{erfc}\left(\frac{y}{2 \sqrt{\nu }}\right)}\,.\label{eq:Fdimd}
\end{align}

The propagator $Q(x, T,t)$ satisfying the equation (\ref{eq:Qeqxit}) with the initial condition $Q(x, T,t=0)=\delta( T-T_f)\delta(x)$ is simply given by
\begin{align}
  Q(x, T,t) = P(x,T_f- T,t)\,,\label{eq:timrevqt}
\end{align}
where we used the time reversibility of Brownian motion. This recovers the expression (\ref{eq:Qdimtm}) displayed in the main text.

\section{Free propagator of the joint position and quadratic area for Brownian motion}\label{app:quad}
In this appendix, we derive the free propagators $P(x,\mathcal{A},t)$ and $Q(x,\mathcal{A},t)$ of the joint position $x$ and quadratic area $\mathcal{A}$ at time $t$ for Brownian motion. We start with the propagator $P(x,\mathcal{A},t)$ in (\ref{eq:Peqxi}) (for $n=2$) and perform a Laplace transform over the variable $\mathcal{A}$ to get
\begin{align}
  \partial_t \tilde P(x,\lambda,t) = D \partial_x^2\tilde P(x,\lambda,t) - \lambda x^2 \tilde P(x,\lambda,t)\,,\label{eq:Ptpq}
\end{align}
where $\tilde P(x,\lambda,t) = \int_0^{\infty}d\mathcal{A} \, P(x,\mathcal{A},t)\,e^{-\lambda \mathcal{A}}$ and where we used $P(x,\mathcal{A}=0,t)=0$ as $\mathcal{A}$ is a positive quantity. By performing an additional Laplace transform over time, one could solve the differential equation (\ref{eq:Ptpq}). We proceed alternatively by first identifying the $x$ dependence of $\tilde P(x,\lambda,t)$ using a path integral method and then using the differential equation (\ref{eq:Ptpq}) to obtain the remaining time dependence. 

The free propagator $P(x,\mathcal{A},t)$ can be written as a path integral
\begin{align}
  P(x,\mathcal{A},t) = \frac{\int_{x(0)=0}^{x(t)=x} \mathcal{D}[x( \tau)] e^{-\frac{1}{4D}\int_0^t d \tau \dot x^2(\tau)}\,\delta\left(\mathcal{A} - \int_0^t d \tau x^2( \tau)\right)}{\int_{x(0)=0} \mathcal{D}[x( \tau)] e^{-\frac{1}{4D}\int_0^t d \tau \dot x^2( \tau)}}\,,\label{eq:pi}
\end{align}
where the numerator contains all the trajectories that start at $0$ and finish at $x$ at time $t$ with a quadratic area $\mathcal{A}$ and the denominator is a normalisation constant that contains all the trajectories from $0$ up to time $t$. The trajectories are weighted by the usual Gaussian weight due to the Gaussian white noise in the equation of motion (\ref{eq:eom}). Performing a Laplace transform with respect to $\mathcal{A}$ we find
\begin{align}
  \tilde P(x,\lambda,t) =\int_0^{\infty}d\mathcal{A} P(x,\mathcal{A},t)\,e^{-\lambda \mathcal{A}}= \frac{\int_{x(0)=0}^{x(t)=x} \mathcal{D}[x(\tau)] e^{-\frac{1}{4D}\int_0^t d \tau \mathcal{L}(x(\tau),\,\dot x( \tau))}\,\delta(x-x(t))}{\int_{x(0)=0} \mathcal{D}[x(\tau)] e^{-\frac{1}{4D}\int_0^t d \tau \dot x^2(\tau)}}\,,\label{eq:pil}
\end{align}
where the Lagrangian is given by
\begin{align}
  \mathcal{L}(x(\tau),\dot x(\tau)) = \dot x^2(\tau) + 4\lambda D \,x^2(\tau)\,.\label{eq:lagrangian}
\end{align}
In fact, one can also interpret the numerator in (\ref{eq:pil}) as the propagator, in imaginary time, of the quantum harmonic oscillator, with mass $m=1/(2D)$ and frequency $\omega = \sqrt{4 D\lambda}$. This propagator can be computed exactly \cite{Feynman65}. One easy way to derive this propagator is to notice that the Lagrangian is quadratic and hence the saddle-point (instanton) method is exact. This means that the functional integral in (\ref{eq:pil}) is completely governed by the classical path. The classical path simply evolves via the Newton's second law which reads 
\begin{subequations}
\begin{align}
\ddot x(\tau)  = 4\lambda D x(\tau) \,,\label{eq:lageq2}
\end{align} 
\end{subequations}
with the boundary conditions $x(0)=0$ and $x(t)=x$. The classical path is therefore
\begin{align}
  x(\tau) = \frac{x\,\sinh(2\sqrt{\lambda D}\, \tau)}{\sinh(2\sqrt{\lambda D}\,t)}\,\,.\label{eq:classicp}
\end{align}
Evaluating the action in the numerator in (\ref{eq:pil}) for the classical path (\ref{eq:classicp}), we find the $x$ dependence of the propagator:
\begin{align}
  \tilde P(x,\lambda,t) = f(t)\,e^{-\frac{x^2}{2}\sqrt{\frac{\lambda}{D}}\,\coth(2\sqrt{\lambda D}t)}\,,\label{eq:xep}
\end{align} 
where $f(t)$ is an unknown time dependence that remains to be found. We now insert the expression (\ref{eq:xep}) into the differential equation (\ref{eq:Ptpq}) to find that $f(t)$ satisfies
\begin{align}
  \sqrt{\lambda D } \,f(t) \coth \left(2 t \sqrt{\lambda D }\right)+\partial_t f(t)=0\,.\label{eq:difff}
\end{align}
Integrating (\ref{eq:difff}), we find that $f(t)$ is given by
\begin{align}
  f(t) = \frac{c(\lambda,D)}{\sqrt{\sinh(2\sqrt{\lambda D}\,t)}}\,, \label{eq:fsol}
\end{align}
where $c(\lambda,D)$ is an integration constant that remains to be determined. Inserting the expression of $f(t)$ (\ref{eq:fsol}) into the propagator (\ref{eq:xep}), we get
\begin{align}
  \tilde P(x,\lambda,t) = \frac{c(\lambda,D)}{\sqrt{\sinh(2\sqrt{\lambda D}\,t)}}\,e^{-\frac{x^2}{2}\sqrt{\frac{\lambda}{D}}\,\coth(2\sqrt{\lambda D}t)}\,.\label{eq:xepf}
\end{align}
From this expression one finds that, as $t\to 0$,
\bea \label{limtzero}
\tilde P(x,\lambda,t=0) = \sqrt{2 \pi} \left( \frac{D}{\lambda}\right)^{1/4} \, c(\lambda,D)\, \delta(x) \;.
\eea
To fix the integration constant $c(\lambda,D)$, we use the initial condition $P(x,\mathcal{A},t=0)=\delta(x)\delta(\mathcal{A})$.  
After a Laplace transform with respect to $\mathcal{A}$, this initial condition in the Laplace space reads $\tilde P(x,\lambda,t=0)=\delta(x)$. 
Comparing this with (\ref{limtzero}) we get $c(\lambda,D)=\lambda^{1/4}/(\sqrt{2\pi} D^{1/4})$. Note that by setting $\lambda=0$ in (\ref{eq:xepf}), we recover the marginal Gaussian distribution for $x$. In addition, by integrating over $x$, we obtain the Laplace transform of the marginal distribution of $\mathcal{A}$
\begin{align}
   \tilde P(\lambda,t) = \frac{1}{\sqrt{\cosh(2\sqrt{\lambda D}t)}}\,.\label{eq:marglamb}
\end{align}
 Performing an inverse Laplace transform in (\ref{eq:xepf}), we find that the propagator is given by
\begin{align}
  P(x,\mathcal{A},t) = \int_\Gamma \frac{d\lambda}{2\pi i} \,e^{\mathcal{A}\lambda}\, \left(\frac{\lambda}{D}\right)^{1/4}\,\frac{e^{-\frac{x^2}{2}\sqrt{\frac{\lambda}{D}}\,\coth(2\sqrt{\lambda D}t)}}{\sqrt{2\pi\,\sinh(2\sqrt{\lambda D}\,t)}}\,,\label{eq:pxil}
\end{align}
where $\Gamma$ is the usual Bromwich contour. In terms of the dimensionless variables $y=x/\sqrt{Dt}$ and $z\mathcal{A}/(Dt^2)$, it takes the scaling form
\begin{align}
   P(x,\mathcal{A},t) = \frac{1}{\sqrt{2\pi D^3\,t^5}}\, \mathcal{G}\left(y=\frac{x}{\sqrt{Dt}},z=\frac{\mathcal{A}}{D t^2}\right)\,,\label{eq:Pdim}
\end{align}
where 
\begin{align}
  \mathcal{G}(y,z) = \int_\Gamma \frac{du}{2\pi i} \, u^{1/4}\,\frac{e^{z\,u-\frac{y^2}{2}\sqrt{u}\,\coth(2\sqrt{u})}}{\sqrt{\sinh(2\sqrt{u})}}\,.\label{eq:Gdim}
\end{align}
The propagator $Q(x,\mathcal{A},t)$ satisfying Eq.~(\ref{eq:Qeqxi}) with the initial condition $Q(x,\mathcal{A},t=0)=\delta(\mathcal{A}-\mathcal{A}_f)\delta(x)$ is simply given by
\begin{align}
  Q(x,\mathcal{A},t) = P(x,\mathcal{A}_f-\mathcal{A},t)\,,
\end{align}
where we used the time reversibility of Brownian motion. This recovers the expression (\ref{eq:Qdim}) displayed in the main text.

\end{document}